\documentclass[%
 amsmath,amssymb,
 aps,
prb
,twocolumn
]{revtex4-2}
\usepackage{epsfig}
\usepackage{graphicx}
\usepackage{dcolumn}
\usepackage{bm}
\usepackage{hyperref}
\usepackage{xcolor}
\usepackage{siunitx}
\usepackage{comment}
\usepackage{cancel}
\usepackage{tensor}
\usepackage{physics}
\usepackage{comment}
\usepackage{siunitx}

\usepackage[normalem]{ulem}
\newcommand{\bR}{\mathbf{R}}

\newcommand{\free}{\mathcal{F}}

\newcommand{\hE}{\hat{\mathcal{E}}}
\newcommand{\E}{\mathcal{H}}

\newcommand{\K}{\mathcal{K}}

\newcommand{\templim}{${T\to 0\;}$}

\newcommand{\bu}{\mathbf{u}}

\newcommand{\dbu}{\mathbf{\dot{u}}}
\newcommand{\bd}{\mathbf{d}}
\newcommand{\brho}{\boldsymbol{\rho}}

\newcommand{\ba}{\mathbf{a}}

\usepackage{mathtools}

\newcommand{\dpsi}{\dot{\psi}}

\newcommand{\bk}{\mathbf{k}}
\newcommand{\hbk}{\hat{\mathbf{k}}}

\newcommand{\bq}{\mathbf{k}}
\newcommand{\newbq}{\mathbf{q}}
\newcommand{\newbbq}{\bar{\newbq}}
\newcommand{\bbq}{\bar{\bq}}

\newcommand{\eig}{\varepsilon}
\newcommand{\loeig}{\varepsilon^{\text{LO}}}
\newcommand{\locorr}{C^{\text{LO}}}

\newcommand{\bomega}{\bar{\omega}}

\newcommand{\bG}{\mathbf{G}}
\DeclareMathAlphabet\mathbfcal{OMS}{cmsy}{b}{n}

\newcommand{\bb}{\mathbf{b}}
\newcommand{\realone}{\mathbb{R}}
\newcommand{\realpos}{\mathbb{R}^+}

\newcommand{\integer}{\mathbb{Z}}

\newcommand{\bz}{\rotatebox[origin=c]{180}{$\Omega$}}
\newcommand{\bza}{\hat{\bz}}

\newcommand{\tC}{\tilde{C}}

\newcommand{\hC}{\hat{C}}

\makeatletter
\newcommand{\subalign}[1]{%
  \vbox{%
    \Let@ \restore@math@cr \default@tag
    \baselineskip\fontdimen10 \scriptfont\tw@
    \advance\baselineskip\fontdimen12 \scriptfont\tw@
    \lineskip\thr@@\fontdimen8 \scriptfont\thr@@
    \lineskiplimit\lineskip
    \ialign{\hfil$\m@th\scriptstyle##$&$\m@th\scriptstyle{}##$&$\m@th\scriptstyle{}##$\hfil\crcr
      #1\crcr
    }%
  }%
}
\makeatother
\usepackage[T1]{fontenc} 
\newcommand{\angstrom}{\text{\normalfont\AA}}

\newcommand{\wk}{${(\bk,\omega)}$}

\newcommand{\tmax}{\mathcal{T}} 
 
\newcommand{\uff}{U_{\textrm{FF}}}
\newcommand{\lofield}{\mathbf{E}^{\text{LO}}}
\newcommand{\loangfreq}{\omega^{\text{LO}}}
\newcommand{\lwaangfreq}{\omega^{\text{LWA}}}
\newcommand{\laangfreq}{\omega^{\text{LA}}}
\newcommand{\toangfreq}{\omega^{\text{TO}}}
\newcommand{\taangfreq}{\omega^{\text{TA}}}

\newcommand{\lolambda}{\lambda^{\text{LO}}}
\newcommand{\brhobk}{\boldsymbol{\rho}_{b,\bk}^{\text{LO}}}

\newcommand{\tv}{\tilde{v}}
\newcommand{\brv}{\breve{v}}

\begin{document}

\title{Breakdown of phonon band theory in MgO}

\author{Gabriele Coiana}
\email{gabriele.coiana17@imperial.ac.uk}
\author{Johannes Lischner}%
\author{Paul Tangney}%
\email{p.tangney@imperial.ac.uk}
\affiliation{%
Department of Materials and Department of Physics, 
Imperial College London, London SW7 2AZ,
United Kingdom.
}%

\date{\today}

\begin{abstract}
We present a series of detailed images of the distribution of 
kinetic energy among frequencies and wavevectors
in the bulk of an MgO crystal as it is heated slowly until it melts.
These spectra, which are Fourier transforms of mass-weighted velocity-velocity 
correlation functions calculated from accurate molecular dynamics (MD) simulations,
provide a valuable perspective on the growth of thermal disorder
in ionic crystals.
We use them to explain why the most striking and rapidly-progressing
departures from a band structure occur among longitudinal optical (LO)
modes, which would be the least
active modes at low temperature ($T$) if phonons did not interact. 
The degradation of the LO band begins, at low $T$, as
an anomalously-large broadening of modes near the center of the Brillouin 
zone (BZ), which gradually spreads towards the BZ boundary.
The LO band all but vanishes before the crystal melts, and
transverse optical (TO) modes' spectral peaks become so broad 
that the TO branches no longer appear band-like.
Acoustic bands remain relatively well defined until
melting of the crystal manifests in the spectra as their sudden disappearance.
We argue that, 
even at high $T$, the long wavelength acoustic (LWA) phonons of an ionic crystal can remain partially immune 
to disorder generated by its LO phonons; whereas,
even at low $T$, its LO phonons can be strongly affected by LWA phonons.
This is because LO displacements average out in much less than the period of an LWA phonon;
whereas during each period of an LO phonon an LWA phonon appears as a quasistatic perturbation 
of the crystal, which warps the LO mode's intrinsic electric field. 
LO phonons are highly sensitive to 
acoustic warping of their intrinsic fields because their 
frequencies depend strongly on them: They cause
the large frequency difference between LO and TO bands known as {\em LO-TO splitting}.
We calculate vibrational spectra from MD trajectories using a method that we
show to be classically exact and therefore applicable, with equal validity, to any solid 
or liquid in any thermal or nonthermal state.
By demonstrating its power and generality we show that it has become possible to go far
beyond the reach of perturbation theories and mean-field theories in the study 
of vibrations in materials. 
\end{abstract}

\maketitle

\section{Introduction}
\label{section:intro}
\begin{center}
\begin{figure*}[!]
    \includegraphics[width=\textwidth]{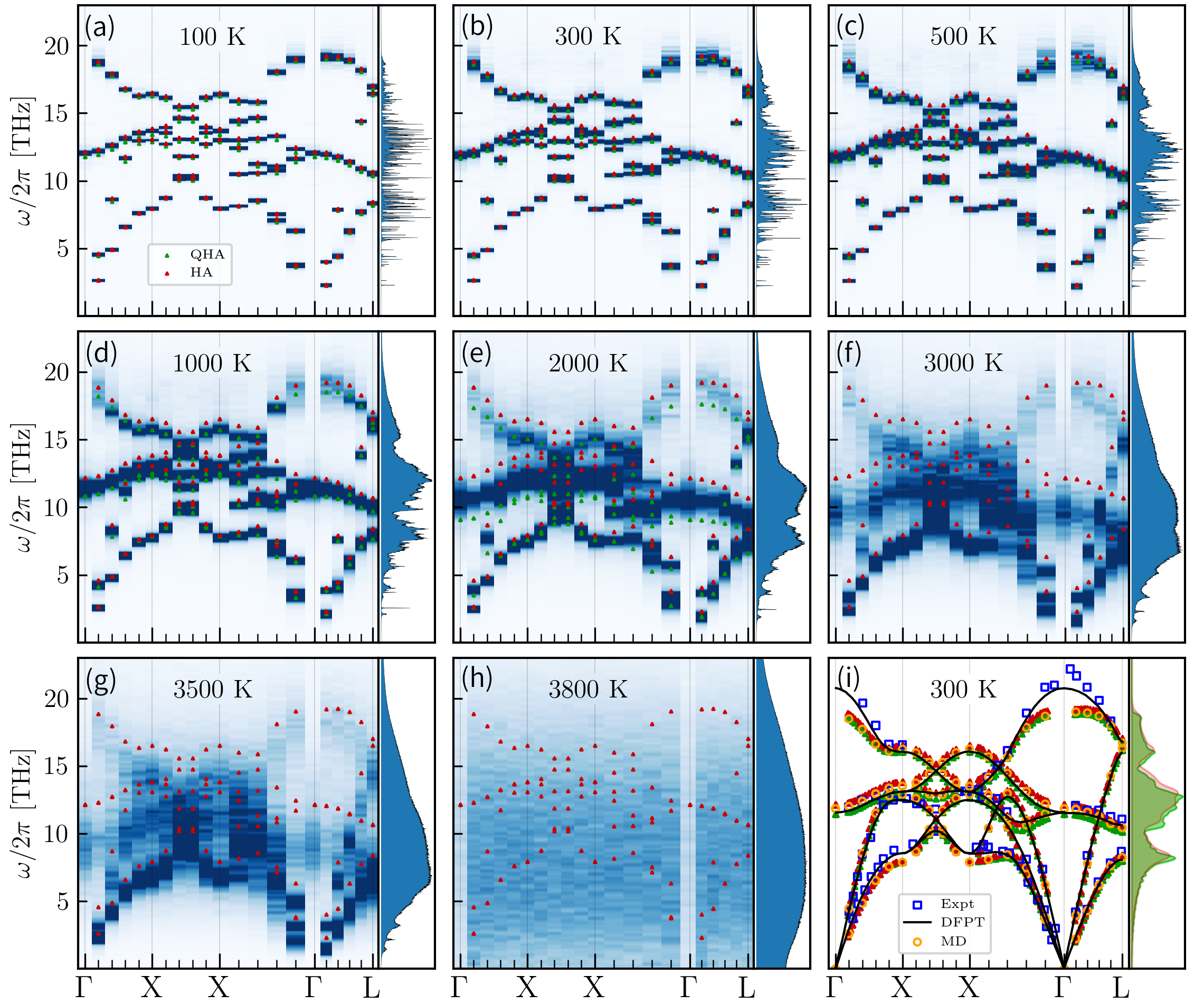}
    \caption{Distribution of kinetic energy of MgO in reciprocal spacetime at a selection of temperatures ($T$) up to and above the melting temperature, $T_m$. 
The values of $T$ are approximate; precise values are provided in Sec.~\ref{section:MD}.
The blue and white background image is data from molecular dynamics simulations using an {\em ab initio}-parameterized
force field, $\uff$: At each $T$ the kinetic energy density in reciprocal spacetime (${\hE^\K(\bk,\omega)}$) was calculated and
normalized to one by dividing it by its integral over the first Brillouin zone. This normalization allowed
the same map between colours and energy density values to be used at each $T$.
The image is pixelated, with the width and height of each pixel equal to the resolutions imposed on our simulations by our finite simulation cell size
and simulation times, respectively. We also project the full spectrum onto the normal mode eigenvectors calculated 
at zero temperature; and we plot the locations of the {\em quasiparticle} peaks as orange symbols in the lower right panel.
The red and green symbols on each plot are the phonon frequencies calculated with ${\uff}$ at zero temperature and 
using the quasiharmonic approximation, respectively. The frequencies measured by Sangster experimentally~\cite{Sangster_1970}
are plotted with blue squares and the black lines were calculated by Karki {\em et. al.} using density functional perturbation theory (DFPT)
and the quasiharmonic approximation~\cite{Karki2000}.
The side plots at each temperature are plots of ${\sum_\bk \hE^\K(\bk,\omega)}$ versus $\omega$, 
with $\omega$ on the vertical axis.
}
    \label{figr1}
\end{figure*}
\end{center}
Band theories are theories of elementary excitations in crystals that are 
derived under the simplifying assumption that the elementary excitations are approximately
independent of one another 
~\cite{wallace,Jones_and_March1,Ashcroft_and_Mermin,cohen_louie_2016}.
For example, the band theory of electrons assumes that excited electronic states of a crystal
are composed of weakly-interacting {\em quasiparticles} (QPs)~\cite{mattuck}, 
which are collective excitations of the electrons in which only one of the electrons plays a prominent
role: A QP can roughly be described as a single electron whose electric field is screened, 
and whose energy and inertia are changed, by its many weak interactions with other electrons. 
When interactions between QPs are strong, they are not mutually independent and each one is
not really a QP, but part of a more complex collective excitation in which multiple electrons play prominent roles~\cite{mattuck}. 
The assumptions underpinning band theory are not valid for such
excitations, and when they cause band theory to break down as an approximation,
the electrons are said to be {\em strongly correlated}.

Similarly, the band theory of phonons assumes that the vibrational energy of a crystal can be approximated
well by a sum of energies of individual {\em phonons}, or {\em phonon quasiparticles}, 
which are lattice waves with well defined frequencies and wavevectors.
Phonon band theory breaks down when strong interactions between phonons with very different frequencies and/or wavevectors
strongly correlate their motions, resulting in motion
that cannot be characterized by a single wavevector and a single frequency, or even by a narrow range
of wavevectors and a narrow range of frequencies.

Although strong electronic correlation has been the subject of intense study
for decades, there have been few fundamental direct studies of strong phononic correlation~\cite{Tangney2006, Zhang2014, breeze, Lu2018, Calandrini2021}.
One reason for this is the so-called {\em terahertz gap}, which is
the region of the electromagnetic spectrum between about ${100\;\text{GHz}}$ and
${\sim 10\;\text{THz}}$ where detectors, and intense continuous wave or pulsed sources, are either
unavailable or not widely available~\cite{thz_gap1,thz_gap2}. 
Another reason is that strong phononic correlation is difficult to 
observe experimentally because measured spectra tend to 
have low resolutions, making it difficult to detect
when, or to what degree, the band theory of weakly interacting phonons has broken down.

In this work we use atomistic simulations to study how the vibrational spectrum of
MgO changes as its temperature ($T$) increases from ${T=100\;\text{K}}$, where
band theory is accurate, to ${T=3800\;\text{K}}$, where the crystal
has melted and the spectrum has lost every semblance of 
a band structure.
At each $T$ we calculate the distribution, ${\hE^\K(\bk,\omega)}$, of kinetic energy in 
{\em \wk-space}, or {\em reciprocal spacetime}, which are the terms we use to refer
to the space of all wavevectors and frequencies. 
We present these spectra in Fig.~\ref{figr1}, and 
analyze them and discuss them in detail in Sec.~\ref{section:results}.

Just as perturbation theories and mean-field approximations (e.g., Hartree-Fock) cannot accurately 
describe strongly-correlated electrons,  mean-field approximations for phonons (so-called {\em
self-consistent phonon approximations}~\cite{hooton_1955,werthamer_1970,wallace,Jones_and_March1}) 
cannot accurately describe strongly-correlated phonons.
Therefore mean-field based methods
of calculating vibrational spectra~\cite{hellman_2013,tadano_2015,bianco_2017,Monacelli_2021,zacharias_2023}
fail when phononic correlation is sufficiently strong, such as at high temperatures ($T$), or in a liquid.
However, the method we use to calculate spectra from our MD simulations is classically exact, which means 
that its accuracy is unaffected by the strength of phononic correlation.

Each distribution ${\hE^\K(\bk,\omega)}$ 
is the Fourier transform (FT), with respect to both space and time, of a mass-weighted velocity-velocity correlation
function (VVCF) calculated from a molecular dynamics (MD) simulation at a different $T$. The MD simulations
were performed with a polarizable-ion force field whose parameters have been 
fit closely to the density functional theory 
(DFT) potential energy surface~\cite{Tangney_Scandolo_2003,Tangney_Scandolo_2009}.

It is common to Fourier transform velocity {\em autocorrelation}
functions (VACFs) with respect to time to produce frequency-resolved spectra, 
such as those along the right-hand vertical edge of each panel in
Fig.~\ref{figr1}.
However, wavevector-resolved spectra are relatively rare: They were
calculated for a one dimensional material in Ref.~\onlinecite{Tangney2006}, 
and more recently they have been calculated for three dimensional crystals using
{\em ab initio} MD~\cite{Sun_2014,Zhang2014,Lu2018}.
However, the computational expense of {\em ab initio} MD severely restricts
the number of frequencies and wavevectors at which ${\hE^\K(\bk,\omega)}$ can be calculated.
Therefore, so far, the resolutions of the spectra calculated by {\em ab initio} MD have been 
too low to see bands.

Force fields whose mathematical forms can mimic the electronic response
to nuclear motion, and whose parameters
are fit to enormous datasets calculated 
{\em ab initio}, provide a very useful balance of speed and accuracy~\cite{Tangney_Scandolo_2002_2, Tangney_Scandolo_2003, behler_2007, 
Csanyi_PRL_2010, Tangney_TiO2_2010, Tangney_Sarsam_2013, Csanyi_Tungsten_2014, Csanyi_2017, machine_learning_2021}. 
As Fig.~\ref{figr1} illustrates, they allow accurate spectra to be calculated with resolutions that are high enough to see a crystal's band structure.
For example, Lahnsteiner and Bokdam~\cite{lahnsteiner} recently used them to calculate detailed spectra at two temperatures 
in order to extract phonon QP frequencies for use within band theory.
Our purpose is very different.

The primary purpose of this work is to study strong phononic 
correlation. To this end, we have undertaken a 
systematic study of the strengthening of phononic correlation, and the consequent
breakdown of band theory, as a crystal is heated. 

We chose MgO for our study because it is a material whose vibrational
properties are important in many contexts, from 
studies of seismic waves travelling through the Earth's lower mantle, 
to its use as a thermal or electrical insulator, 
as a substrate for growing superconducting or ferroelectric perovskites, 
and in countless other important applications\cite{Karki2000,Ghose_2006,Tangney_Scandolo_2009,breeze,Dekura2013,Kimura_2017}.
As well as being a technologically-important material, MgO is one of the 
simplest oxides: It is a strongly ionic insulator with the same cubic
crystal structure as rocksalt. For these reasons, it plays a similar representative
role for oxides to that played by silicon for semiconductors: It is 
often the simplest setting in which properties or phenomena that are 
common to many oxides can be investigated.
Several experimental and computational studies of phonon-phonon interactions in MgO and
similar materials have
recently been published~\cite{breeze, Lazzeri_2018, Giura2019, Calandrini2021, Togo_2022}; therefore it is
a natural and obvious starting point for investigating strong
phononic correlation in oxides. 

Calculations of detailed accurate
spectra like those presented in Fig.~\ref{figr1} and by 
Lahnsteiner and Bokdam have been possible for a decade or more.
One reason for their rarity 
may be that it is not commonly known that, within classical physics, the FT
of the VVCF, ${\hE^\K(\bk,\omega)}$, is \emph{exactly}
the distribution of kinetic energy in reciprocal spacetime:
To our knowledge, all existing derivations and discussions of the theory 
rely on two simplifying assumptions: They assume that the crystal is at thermal equilibrium
and that $T$ is low enough for the vibrational spectrum to be a band structure~\cite{dove_1993,vanderbilt_ice}.
After making these assumptions, the {\em equipartition theorem} is usually invoked to
relate ${\hE^\K(\bk,\omega)}$ to the vibrational density of states (VDOS)\cite{dove_1993,vanderbilt_ice}.

Lahnsteiner and Bokdam are among those who state that ${\hE^\K(\bk,\omega)}$ (our notation)
is the wavevector-resolved VDOS, and they support this statement by citing Refs.~\onlinecite{Ladd_1986,Sun_2010,Zhang2014,Sun_2014}.
This interpretation of ${\hE^\K(\bk,\omega)}$, which can also be 
found in many other works~\cite{dove_1993,vanderbilt_ice,Meyer_2011,Kirchner_2013,tuckerman,meunier},
is correct in the limit \templim and under the simplifying assumptions that lead to band theory. 
It is also the appropriate interpretation of existing theory, because
derivations such as those of Dove~\cite{dove_1993} and Lee {\em et al.}~\cite{vanderbilt_ice}
only provide a clear physical interpretation of ${\hE^\K(\bk,\omega)}$ at thermal equilibrium in the \templim limit.

However, it is shown in App.~\ref{section:string} and Ref.~\onlinecite{theory_paper} that Fourier transforming the VVCF is a much more 
powerful and general method than it is currently believed to be:
It is shown that ${\hE^\K(\bk,\omega)}$ is exactly the
distribution of kinetic energy in reciprocal spacetime. 
Appendix~\ref{section:string} contains an illustrative proof of this result, for the
case of a vibrating string.
Therefore each of the spectra presented in Fig.~\ref{figr1} is {\em exactly}
the distribution, among points \wk~in reciprocal spacetime,
of the classical kinetic energy of the MD simulation from which it was calculated.

Furthermore, proving that ${\hE^\K(\bk,\omega)}$ is the kinetic energy
distribution does not require any assumptions to be made about
the statistical state or the structure of the material. 
It is a result that applies, with equal theoretical validity, to any crystalline 
or amorphous material, in any stationary or nonstationary state, regardless
of the strength of the correlation between different vibrations and waves.
This means that its range of possible applications is vast. 

For example, it could be used to calculate the spectra of solids or liquids while they are
being resonantly excited by THz radiation; or to study how spectra change
during order-disorder phase transitions; or to investigate the relationships between
structure, energetics, and diffusion in liquids.
Using it may deepen our understanding of strongly correlated electronic systems, which 
are much harder to study computationally, by
providing insight into aspects of strong correlation that are
common to both phononic and electronic systems.
It can also be used to assess the accuracies of approximate methods, such
those based on perturbation theory, self-consistent mean-field methods,  
and methods that assume thermal equilibrium.
As it is the only method that provides the exact classical spectrum, it is the method against which
the accuracies of all other methods should be judged.
A secondary purpose of this work is to demonstrate the power of this 
theoretical result and computational method.

By applying it to MgO we uncover a simple, strong, and general non-resonant mechanism by which the highest frequency
optical modes of an ionic crystal can be disrupted by the lowest frequency acoustic
modes, leading to the disintegration, or {\em melting}, of the optical bands.
We call this mechanism {\em acoustic warping of optical phonon fields}.
We also explain why, despite this mechanism causing longitudinal optical (LO)
bands to melt, the acoustic bands responsible for their melting 
remain intact: It is because acoustic bands are {\em adiabatically decoupled}
from LO phonons in the same way that heavy nuclei are adiabatically decoupled from 
electrons, despite pushing the electron density around as they move.
Among many other analyses of the results presented in Fig.~\ref{figr1}, we
compare them to a second set of spectra (Fig.~\ref{fig:fig2}) which were calculated
from MD simulations performed at the \templim~density. This allows us to discover
which $T$-induced changes to the spectrum can be explained by thermal expansion and which cannot.
It also provides insight into the strengths and limitations 
of the {\em quasiharmonic approximation}~\cite{cowley_qha_1963,dove_1993,Karki2000}.

In the next section we discuss the effects of $T$ on phonon band structures
in general terms. 

In Sec.~\ref{section:theory} we explain our notation and
some aspects of phonon theory that can be different when phonons are strongly correlated
than they are at low $T$ where perturbation theories are applicable.
For example, at high $T$ each mode is not Lorentzian, in general, and so the 
Lorentzian width is not a good measure of the degree to which $T$
has broadened it.

In Sec.~\ref{section:methods} we explain how we performed our simulations.

We begin Sec.~\ref{section:results} by discussing the most important limitations
of our simulations. Then we begin discussing and numerically-analysing the spectra presented
in Fig.~\ref{figr1} and Fig.~\ref{fig:fig2}. We focus our discussions and analyses 
on the most important and striking features of these spectra, and we include
an explanation of the acoustic warping mechanism.

We summarize our conclusions in Sec.~\ref{section:conclusions}.

\section{Qualitative effects of temperature on band structures}
\label{section:T_effects}
In this section we discuss the qualitative effects of temperature on vibrational spectra.
We begin with an illustration of a pertinent mathematical point.

\begin{figure}[!]
    \includegraphics[width=0.9\columnwidth]{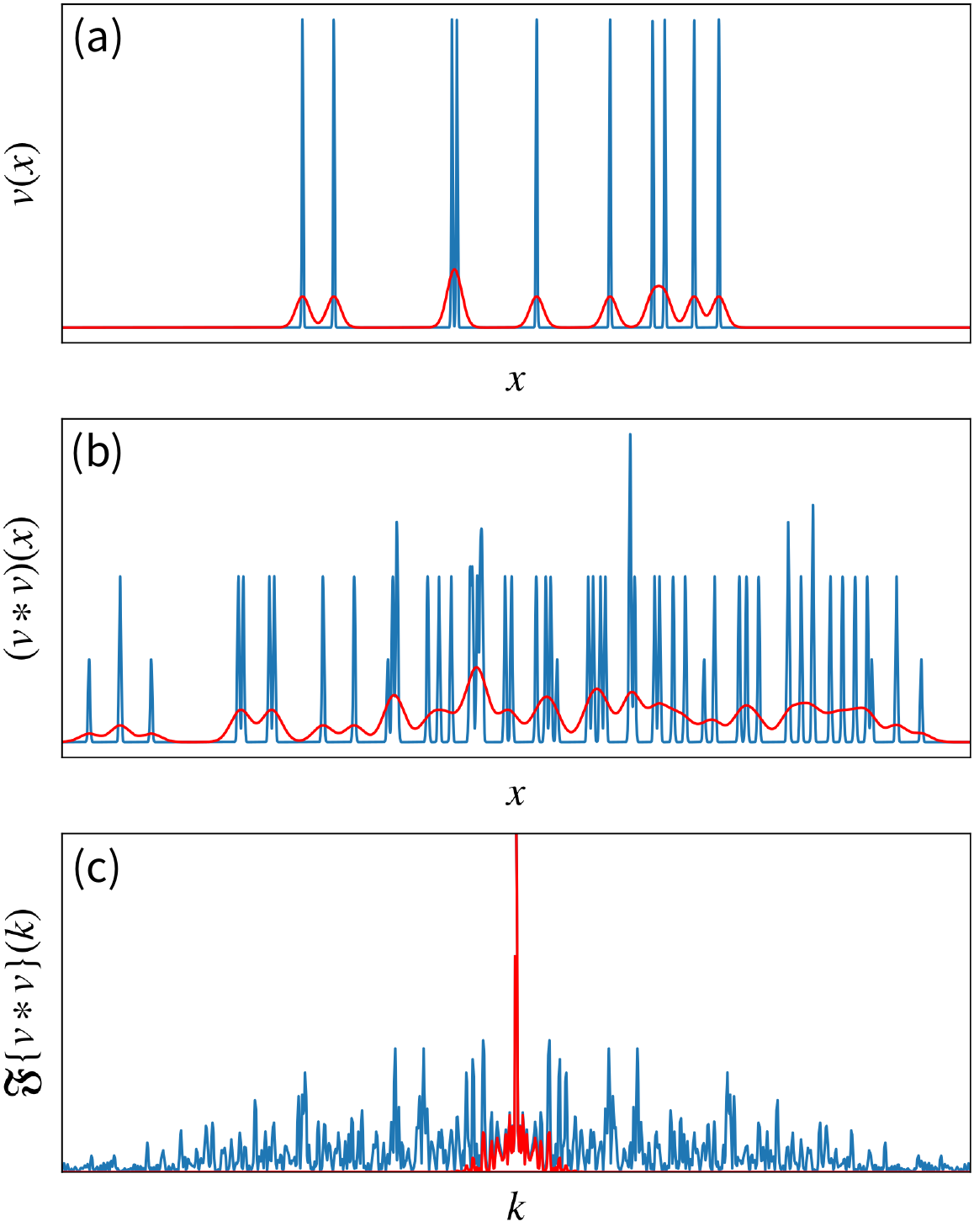}
    \caption{(a) Each of the plotted curves is the sum of an array of Gaussians
at the same randomly-chosen positions. The only difference between the curves is that
the widths of the Gaussians contributing to 
red curve are ten times larger than those contributing to the blue curve;
(b) The convolutions of each curve plotted in (a) with itself;
(c) The Fourier transforms of the convolutions plotted in (b). 
}
    \label{fig:comb}
\end{figure}
Figure~\ref{fig:comb}(a) is a plot of two curves, each of which is the sum of 
ten randomly-positioned Gaussians.
The only difference
between the two curves is that the width of the Gaussians contributing to the red curve is a factor
of ten larger than those contributing to the blue curve. 
The convolution of the red curve with itself and the convolution 
of the blue curve with itself are plotted in Fig.~\ref{fig:comb}(b), 
and the Fourier transforms of these convolutions are plotted in Fig.~\ref{fig:comb}(c).

It is well known that the more localized a smooth function is, the more delocalized 
its Fourier transform is~\cite{strichartz}. These plots illustrate that the Fourier transform
of the convolution of a smooth function with itself is more delocalized
when the function is localized and vice-versa.

${\hE^\K(\bk,\omega)}$ is the FT of the VVCF and the VVCF is 
the correlation of the spacetime distribution of $\sqrt{\text{mass}}$-weighted atomic velocities with itself.
The spacetime distribution of the atoms' kinetic energy is more localized/delocalized
when the spacetime distribution of their velocities are more localized/delocalized.
Therefore Fig.~\ref{fig:comb} illustrates
the fact that as kinetic energy becomes more localized in real spacetime
it becomes more delocalized in reciprocal spacetime and vice versa.
With this in mind, we now discuss the qualitative
effects of temperature on vibrations in crystals.

\subsection{Normal mode vibrations,  phonon quasiparticles, and strong phononic correlation}
Classically, and in the \templim limit, the term {\em phonon} refers to 
an oscillation of one of a crystal's normal modes of vibration, which are 
standing waves of the lattice that all of the crystal's atoms participate in.
Each normal mode is characterized by a single frequency ($\omega$) and
a single wavevector ($\bk$)~\cite{theory_paper, wallace, Ashcroft_and_Mermin}. 
Therefore the energy of each normal mode vibration (NMV)
can be regarded as localized at a point 
\wk~in reciprocal spacetime.

The amplitudes of NMVs vanish in the \templim~limit, making
them perfectly harmonic and mutually noninteracting.
Therefore the energy of each one is delocalized in spacetime.
It is delocalized in space because every atom in the crystal participates in it and shares its energy.
It is `delocalized' in time in the sense that it is not transient: NMVs last forever in the \templim~limit 
because, without interactions, they cannot dissipate their energies; and they
cannot disperse because each one is characterized by a single frequency 
and a single wavevector.

As well as being the limit in which kinetic energy is delocalized
in spacetime and localized at a point in reciprocal spacetime, the \templim~limit is
the limit in which atoms are strongly correlated and the limit in which phonons are uncorrelated:
Each contribution to the atoms' kinetic energy is a collective motion of all atoms, whereas each phonon 
possesses its own kinetic energy, which is independent of the energies of other phonons.

At finite $T$, phonons' amplitudes are finite, and they interact to some degree.
When they interact, their motions become correlated and each one is no longer characterized by a single
frequency and wavevector: 
it has become a superposition of NMVs with different frequencies and wavevectors. 
This interaction-induced mixing of frequencies and wavevectors is analogous to what happens when a pair of
harmonic oscillators with frequencies $\omega_1$ and $\omega_2$ are coupled:
The motion of each one becomes characterized by both frequencies; or equivalently, each one becomes an oscillation
at frequency ${\frac{1}{2}(\omega_1+\omega_2)}$ whose amplitude is modulated
by an oscillation at frequency ${\frac{1}{2}|\omega_1-\omega_2|}$.
When many oscillators are coupled strongly, the motion
of each one becomes a superposition of oscillations at many different frequencies.
Similarly, interactions cause each phonon's energy
to be distributed among the frequencies and wavevectors of all 
lattice waves contributing to its motion.

\subsubsection{Phonon quasiparticles}
At low finite $T$ each phonon is still mostly composed of a single NMV, but it contains small
contributions to its motion from the other NMVs with which it interacts~\cite{wallace}. 
As $T$ increases and interactions strengthen the fraction of its kinetic energy
contributed by the original NMV decreases and the fractions contributed by others increase. 
Therefore, as $T$ increases from the \templim~limit, each NMV's kinetic energy spreads out from the point \wk~at which it was
localized and becomes distributed among the frequencies
and wavevectors of all of the NMVs with which it interacts.

At low $T$ the vibrational spectrum remains sharply peaked 
near the frequencies and wavevectors of the normal modes, but the
peaks have finite widths. This means that each phonon contributing
to each peak is a QP, i.e.,  a
superposition of NMVs with very similar wavevectors and frequencies.
Phonon QPs can dissipate their energies and disperse: 
their average lifetime is inversely proportional to the width of their spectral
peak~\cite{wallace,quasiparticles}. Therefore, at low $T$ they are still reasonably long-lived excitations
with reasonably well-defined frequencies and wavevectors, and one can
think of each one as an NMV that is dressed by its 
interactions with other NMVs. 

Interactions make the average frequency and wavevector
of the QPs $T$-dependent, and cause their 
spatial coherence (size) and temporal coherence (lifetime)
to reduce as $T$ increases.
The reduction in their sizes means that they are
no longer collective motions of all atoms in the crystal. Therefore they
are no longer standing waves, but travelling wave packets.

The kinetic energy spectrum, ${\hE^\K(\bk,\omega)}$, is called
a {\em band structure} at low $T$ because the set of points at which 
there are spectral peaks forms a set of three dimensional surfaces,
or {\em bands}, in four-dimensional reciprocal spacetime.
As each phonon becomes a mix of NMVs, these
spectral peaks broaden and move, which causes the bands to blur, lose definition, 
and shift in frequency. 

In quantum mechanics, the energies of NMVs and QPs are quantized and 
it is the quanta that are known as phonons. However, because most of our simulations and 
analyses are classical, we use the term phonon to refer to NMVs, in the \templim~limit, and to phonon QPs, at finite $T$.
A classical phonon is simply a vibration of the crystal whose wavevector and
frequency are reasonably well defined, meaning that the distribution
of its energy in reciprocal spacetime is sharply peaked, and most
of its energy is localized in a small neighbourhood of its peak.

\subsubsection{Strong phononic correlation}
As $T$ increases further, sooner or later the $T$-induced changes to the spectrum become
more complex than a gradual shifting and symmetric broadening
of QP peaks. Peaks may broaden so much that they vanish; one peak may become two 
peaks at very different frequencies and wavevectors; or the spectrum may deviate from a superposition
of well-defined QP peaks in other ways, depending on the natures and strengths
of the interactions.
When the spectrum can no longer be approximated by a sum of
reasonably-localized QP peaks, the quasiparticle approximation has broken down
and the system can be regarded as strongly correlated.

When interactions between phonons are strong enough
that the QP approximation has broken down, band theory also breaks down,
because kinetic energy is so delocalized in reciprocal
spacetime that it no longer forms bands.

For example, the energy of each oscillation in a liquid is localized in spacetime
and delocalized in reciprocal spacetime.
Oscillations are localized in space because spatial correlations are very short-ranged, which implies that 
only a small cluster of atoms participates in each one. They are localized in time because their strong coupling to 
other oscillatory and translational motions makes them transient: they have short lifetimes because they
quickly disperse and/or dissipate their energies.
Their short lifetimes can also be viewed as them morphing into other forms of motion, namely, 
diffusive motion or oscillations of different frequencies.
For example, we can view the short lifetime of an oscillation of an atom as a consequence of 
the potential well in which it oscillates changing shape as neighbouring atoms move.

The eventual failure of band theory as $T$ increases
is inevitable for every material 
because temperature generates disorder 
and because each point on a band
represents a phonon with a different characteristic pattern of atomic displacements, 
or {\em eigenvector}, and with periodicities ${\lambda\equiv 2\pi/\abs{\bk}}$ and
${\tau\equiv 2\pi/\omega}$ in space and time, respectively. 
When such a vibration exists, it causes correlation between the velocity of an atom 
at time $t$ and the velocity of another atom, which is displaced from it
by $\lambda$ in the direction of ${\bk}$, at time ${t+\tau}$. 
Therefore, when there is a $T$-induced reduction of the correlation length
to less than  $\lambda$, or of the correlation time to less than $\tau$, 
the spectral intensity at point \wk~almost vanishes. 
When a particular mode or band loses all or most of its intensity in this way, we say that
it has melted.

Band melting occurs gradually at most temperatures;
and, as Fig.~\ref{figr1} illustrates, it occurs at different rates for different bands,
because correlation lengths and times are different for motions
along different eigenvectors, in general. 
The rate at which each band melts is determined by the
natures and strengths of the interactions between the band's phonons and other phonons.
However, as Fig.~\ref{figr1}(g) and Fig.~\ref{figr1}(h) illustrate,
when an acoustic band melts suddenly and completely, it means that the crystal has become structurally unstable and has undergone
a phase transition. In some crystals, such as MgO, this does not occur until it becomes
a liquid at the melting temperature of the crystal, $T_m$. In others, such as
ferroelectric BaTiO$_3$~\cite{Cochran_1960,stern_2004,Hlinka_2008,Gu_2021}, there are also one or more transitions between
crystalline phases at ${T}$'s lower than ${T_m}$.

\section{Elements of the theory of interacting phonons}
\label{section:theory}
In this section we present some elements of phonon theory
that we will use in this work. A more complete account of
this theory can be found in Ref.~\cite{theory_paper} and elsewhere
in the literature~\cite{Jones_and_March1, wallace, Ashcroft_and_Mermin}.

\subsection{Notation and definitions of key quantities}
\label{section:notation}
\subsubsection{Structure of the crystal}
We use $N$ to denote the number of atoms in each primitive cell, which is two in this case, 
and we denote the number of primitive cells in the crystal's bulk by $N_c$.
We denote the primitive lattice vectors by ${\{\ba_1,\ba_2,\ba_3\}}$
and we identify primitive cells by their positions, $\bR$, relative to an origin in
the bulk of the crystal. These positions are lattice vectors, i.e., 
${\bR=\sum_{\alpha=1}^3 R^\alpha\ba_\alpha}$, where ${R^1, R^2, R^3\in\integer}$.

We use the compound index ${\bR j}$ to identify the ${j^\text{th}}$ atom in primitive cell
${\bR}$ and we use ${\bR j \alpha}$ to identify its ${\alpha^\text{th}}$ lattice
coordinate. We denote its displacement, at time $t$, from its ${T\to 0}$ equilibrium position 
as
${\bu^{\bR j}(t)\equiv \sum_{\alpha}u^{\bR j \alpha}(t)\,\ba_\alpha}$.
However, it is not very convenient to 
use $N$ vectors ${\bu^{\bR j}\in\realone^3}$ to
specify the internal structure of each primitive cell.
Instead, we specify it with a single vector ${\ket{\psi_\bR}\equiv \sum_{j\alpha} \psi^{\bR j \alpha}\ket{j \alpha}\in\realone^{3N}}$, 
where ${\psi^{\bR j \alpha}\equiv\sqrt{m_j}u^{\bR j \alpha}}$, $m_j$ is the mass
of the ${j^\text{th}}$ atom, and the set ${\{\ket{j\alpha}: \alpha\in\{1,2,3\}, j\in\{1,\cdots,N\}, \braket{j\alpha}{k \beta}=\delta_{jk}\delta_{\alpha\beta}\}}$
is a complete orthonormal basis of ${\realone^{3N}}$.

For simplicity, and despite its ${\sqrt{\text{mass}}}$ weighting, 
we will often refer to the vector ${\ket{\psi_\bR(t)}}$ and its time 
derivative ${\ket{\dpsi_\bR(t)}}$ as the {\em displacement} and {\em velocity}, respectively, 
of primitive cell $\bR$.
The kinetic energy of cell $\bR$ is 
${\frac{1}{2}\braket{\dpsi_{\bR}}=\frac{1}{2}\sum_j m_j \abs{\dbu^{\bR j}}^2}$.

\subsubsection{Correlation functions and their Fourier transforms}
\label{section:theory_spectra}
The mass-weighted velocity-velocity correlation function (VVCF) is
\begin{align}
C(\bR,t) \equiv \expval{\braket{\dpsi_{\bR_0}(t_0)}{\dpsi_{\bR_0+\bR}(t_0+t)}}_{\bR_0,t_0}
\label{eqn:correlation}
\end{align}
where the average is performed over bulk cells,  ${\bR_0}$,
and over times, $t_0$.
It is shown in Ref.~\cite{theory_paper} that the average kinetic
energy per bulk unit cell is given by
\begin{align}
\frac{\expval{\K}}{N_c} \equiv \sum_{\bk}\sum_{\omega>0} \hE^\K(\bk,\omega),
\label{eqn:kinetic}
\end{align}
where $\K$ is the total kinetic energy of all bulk cells; 
${\expval{\K}}$ is its time average;
and
${\hE^\K(\bk,\omega)}$ is the
discrete Fourier transform of ${C(\bR,t)}$ with respect to both $\bR$ and $t$.
${\hE^\K(\bk,\omega)}$ is the distribution, in reciprocal spacetime, of the
time-average of the crystal's kinetic energy per bulk primitive cell.
As discussed in Sec.~\ref{section:intro}, Eq.~\ref{eqn:kinetic} is an exact expression,
which is proved for the case of a vibrating string in Appendix~\ref{section:string}, 
and for a crystal in Ref.~\cite{theory_paper}.
Theoretically, it is no less valid to apply it to a nonequilibrium liquid 
than it is to apply it to a low temperature crystal.

The first sum in Eq.~\ref{eqn:kinetic} is over the set $\bza$ of all wavevectors $\bk$ within
the first Brillouin zone, $\bz$, that
are compatible with the boundary conditions of the crystal.
Within our treatment of the theory, 
wavevectors that differ
by a reciprocal lattice vector, $\bG$, are regarded as equivalent
to the same wavevector in the first Brillouin zone, $\bz$.
If we wanted to distinguish between them, the right hand side of 
Eq.~\ref{eqn:kinetic} would have the form 
${\sum_{\bk}\sum_{\omega>0}\left(\sum_\bG F(\bk+\bG,\omega)\right)}$, 
where $F$ is a function whose domain is the set of all wavevectors.
However we are choosing to define ${\hE^\K(\bk,\omega)\equiv \sum_\bG F(\bk+\bG,\omega)}$
and to restrict our attention to the finite set of wavevectors $\bza$.
This means that we do not explicitly distinguish between
so-called {\em Umklapp} phonon interactions and normal interactions.
This distinction is commonly made when treating interactions as scattering events, but 
in this work we emphasize the wave natures of phonons: Both in our classical simulations 
and, when $T$ is high enough, in a real crystal, phonons 
exchange energy with one another quasi-continuously.

The second sum in Eq.~\ref{eqn:kinetic} is over all possible frequencies.
This set is countable because if the total time for which a crystal is 
observed or simulated is $\tmax$, 
complete oscillations whose periods are longer than $\tmax$
are not observed or simulated. Therefore oscillations with 
frequencies smaller than ${2\pi/\tmax}$ cannot contribute
to ${C(\bR,t)}$. Furthermore, any two frequencies that do contribute
to ${C(\bR,t)}$ are only {\em observably} different if they differ by more 
than ${2\pi/\tmax}$~\cite{theory_paper}. Therefore 
${\hE^\K(\bk,\omega)}$ is only defined for ${\omega \equiv 2\pi m/\tmax}$, 
where ${m}$ is a nonnegative integer.

In a crystal, the \templim limit of  ${\hE^\K(\bk,\omega)}$ is a band structure. Therefore, 
when studying the breakdown of band theory, it is 
useful to decompose it into contributions from motions along each of the crystal's
normal mode eigenvectors.
In Ref.~\cite{theory_paper} and many textbooks~\cite{wallace, Ashcroft_and_Mermin, Jones_and_March1}
it is shown that, deep within the bulk of a large crystal, 
each normal mode is associated with a particular wavevector, $\bk\in\bza$,
and can be labelled by ${\bk\mu}$, where $\mu\in\{1,\cdots,3N\}$ is the
{\em band index}. 
At each wavevector $\bk$ the dynamical matrix has ${3N}$ real eigenvectors, ${\ket{\eig_{\bk\mu}}\in\realone^{3N}}$, 
where ${\mu\in\{1,\cdots,3N\}}$ is a {\em band index}.
We choose them to have unit norms and refer to them as the {\em cell eigenvectors}. 
At each ${\bk}$, the set of cell eigenvectors, ${\left \{\ket{\eig_{\bk\mu}}\right\}_{\mu=1}^{3N}}$,
is orthonormal (${\braket{\eig_{\bk\mu}}{\eig_{\bk\nu}}=\delta_{\mu\nu}}$)
and is a complete basis of ${\realone^{3N}}$.
The cell eigenvectors are ${\sqrt{\text{mass}}}$-weighted
{\em polarization vectors}~\cite{Ashcroft_and_Mermin, Jones_and_March1}, 
but polarization vectors are not mutually orthogonal when there
is more than one atomic species.

For any wavevector ${\bk}$, we can express the identity in $\realone^{3N}$ as ${\sum_{\mu=1}^{3N}\dyad{\eig_{\bk\mu}}}$
and insert it into Eq.~\ref{eqn:correlation}, to define the set of
{\em mode-projected correlation functions}
at wavevector ${\bk}$,
\begin{align}
&C_{\bk\mu}(\bR,t)
\equiv  \expval{\braket{\dpsi_{\bR_0}(t_0)}{\eig_{\bk\mu}}\braket{\eig_{\bk\mu}}{\dpsi_{\bR_0+\bR}(t_0+t)}}_{\bR_0,t_0}.
\nonumber
\end{align}
From this definition, and Eqs.~\ref{eqn:correlation} and~\ref{eqn:kinetic}, it
follows that ${C(\bR,t) = \sum_{\bk\mu} C_{\bk\mu}(\bR,t)}$ and
${\hE^\K(\bk,\omega) = \sum_{\bk\mu} \hE^\K_{\mu}(\bk,\omega)}$, where
${\hE^\K_{\mu}(\bk,\omega)}$ denotes the discrete Fourier transform 
of ${C_{\bk\mu}(\bR,t)}$ with respect to ${\bR}$ and $t$.
In Sec.~\ref{section:results} we will pay particular attention to the LO mode and 
we will denote its cell eigenvector and mode-projected correlation
function at $\bk$ by ${\ket{\loeig_\bk}}$ and ${\locorr_\bk(\bR,t)}$, respectively.

When studying individual modes, we will make use of the {\em normalized mode spectrum} or {\em mode distribution}
of mode ${\bk\mu}$, defined as 
\begin{align}
f_{\bk\mu}(\omega) \equiv \frac{\hE^\K_{\mu}(\bk,\omega)}{\sum_{\omega}\hE^\K_{\mu}(\bk,\omega)}.
\label{eqn:mode_spectrum}
\end{align}
This function is the distribution among frequencies $\omega$
of the energy of oscillatory motion in the bulk of the crystal
along the eigenvector of normal mode $\bk\mu$.

When studying the spectrum as a whole, calculating ${\hE^\K(\bk,\omega)}$ from ${C(\bR,t)}$ is 
equivalent to calculating ${\hE_{\mu}^\K(\bk,\omega)}$ from
${C_{\bk\mu}(\bR,t)}$ for each mode ${\bk\mu}$ and summing over all modes.
Therefore, the spectra plotted in Fig.~\ref{figr1} are plots of
\begin{align}
f(\bk,\omega)\equiv
\frac{\hE^\K(\bk,\omega)}{\sum_{\bk\omega}\hE^{\K}(\bk,\omega)}
=
\frac{\sum_{\mu} \hE_{\mu}^\K(\bk,\omega)}{\sum_{\bk\omega} \sum_\mu \hE_{\mu}^\K(\bk,\omega)}.
\label{eqn:full_spectrum}
\end{align}
At thermal equilibrium, the equipartition theorem states that the expectation value of the kinetic
energy of each mode ${\bk\mu}$ is ${\frac{1}{2}k_B T}$. Therefore, if the spectra plotted in Fig.~\ref{figr1} 
were converged fully, the denominators
of both expressions in Eq.~\ref{eqn:full_spectrum} would be ${\frac{3}{2}N k_B T}$.
However, regardless of whether or not ${f(\bk,\omega)}$ is converged with respect to
simulation time or supercell size, it is the exact distribution,
among wavevectors and frequencies, of the average kinetic energy per primitive
cell of the trajectory from which the VVCF was calculated.

\subsection{Energy expanded in mode coordinates}
The displacement of the crystal from equilibrium can be expessed
as a sum of displacements along its ${3 N N_c}$ normal mode eigenvectors.
We denote the frequency of mode ${\bk\mu}$ by ${\omega_{\bk\mu}}$, 
and the projection of the crystal's displacement
from equilibrium at time $t$ onto its normalized eigenvector by
${Q_{\bk\mu}(t)}$, which we refer to as its {\em mode coordinate}.
The total energy of the crystal, as a 
function of the mode coordinates and their time derivatives, can be expressed as 
\begin{align}
\E = \E_0+\K+\E_2 + \E_3 + \E_4 + \E_5 + \cdots,
\label{eqn:energy_expansion}
\end{align}
where ${\E_0}$ is the potential energy of the static lattice 
in the ${T\to 0}$ limit (excluding zero point energy);
$\E_1$ vanishes because the first partial derivatives
of the potential energy vanish at equilibrium;
the kinetic energy is ${\K= \frac{1}{2}\sum_{\bk\mu}\abs{\dot{Q}_{\bk\mu}}^2}$;
the harmonic term of the potential energy is
${\E_2 = \frac{1}{2}\sum_{\bk\mu}\omega_{\bk\mu}^2 \abs{Q_{\bk\mu}}^2}$, 
and $\E_p$, where ${p>2}$, denotes an anharmonic term of order ${Q^p}$.

\subsection{Difference between the kinetic energy distribution and the VDOS}
As discussed in Sec.~\ref{section:intro}, ${\hE^\K(\bk,\omega)}$ is usually interpreted
as being proportional to the VDOS at low $T$~\cite{Ladd_1986,dove_1993,vanderbilt_ice,Sun_2010,Meyer_2011,Kirchner_2013,tuckerman,meunier}.
However, this interpretation is only valid under the simplifying assumptions that 
there are as many vibrational states as there are degrees of freedom and that the crystal
is at thermal equilibrium. When these assumptions are not valid,
the kinetic energy distribution is a very different quantity to the VDOS because
there is kinetic energy at \wk~if one or more waves exist with wavevector $\bk$ and frequency $\omega$;
and because the most general and rigorous treatments of phonon theory do not
place any restrictions on which waves may exist at finite $T$.


Therefore, let us assume that one of the defining characteristics of each `state' that contributes 
to the VDOS is that it is independent, or approximately independent in the temperature range of interest,
of the amount of kinetic energy that `occupies' it.
Then, the only viable definition of the set of all states is this one: at every point \wk, there are exactly as 
many occupiable vibrational states as there are degrees of freedom in a primitive unit cell.

For example, in our MD simulations, there could be kinetic energy at any point 
on a four-dimensional lattice in reciprocal spacetime whose
lattice spacing along the frequency axis is inversely proportional to the total simulation
time and whose lattice spacings along the three wavevector axes are inversely proportional
to the linear dimensions of the simulation supercell.
Therefore the VDOS of the simulated system is a uniform distribution with exactly ${3N=6}$ states
per point \wk~of the lattice that our simulation samples.

To illustrate that this is the case, note that each spectrum in Fig.~\ref{figr1} is pixelated. The 2-d lattice consisting of points at the centers
of the pixels is a 2-d representation of the 4-d VDOS. Therefore there are six states at the center of every pixel, including
at the centers of pixels that are white. 
Pixels are white when none of the states at their centers are `occupied' by kinetic energy. However, the fact that they are not occupied does not mean
that they are not occupiable, and some of them can be seen to turn blue at high $T$ (e.g., when the crystal melts), indicating that 
they have become `occupied' by some of the kinetic energy.

\subsection{Independent-phonon approximations}
In the bottom right panel of Fig.~\ref{figr1} we compare
the ${T = \SI{300}{\kelvin}}$ band structures measured
experimentally~\cite{Sangster_1970} with those that we have calculated
using three different (quasi)independent-phonon approximations; namely, 
the harmonic approximation (HA), the quasiharmonic approximation (QHA), and the quasiparticle approximation (QPA).

In the HA, all cubic and higher-order anharmonic terms 
in the potential energy are discarded, 
which is tantamount to assuming that there
is no interaction between different modes, and
the Helmholtz free energy, ${\free(V,T)}$, can be approximated as~\cite{wallace}
\begin{align}
    \label{eqF}
    \free_H(V,T) = \E_0(V) 
&+ \sum_{\mathbf{k} \mu} \bigg[\frac{1}{2}\hbar\omega_{\mathbf{k} \mu}(V) 
\nonumber\\
 + &k_B T \ln \left( 1 - e^{-\hbar \omega_{\mathbf{k}\mu}(V)/ k_B T} \right)\bigg],
\end{align}
where $k_B$ is the Boltzmann constant, and the `$H$' 
subscript on ${\free_H}$ 
indicates that it is the free energy within the {\em harmonic} approximation, ${\E\approx \E_0 + \K + \E_2}$.
Notice that $\free_H$ only depends on $T$ via the
second term in the summation over modes, but that it
depends on $V$ via the cohesive energy, ${\E_0(V)}$, 
and the \templim~mode frequencies, ${\omega_{\bk\mu}(V)}$.

When the third term vanishes in the \templim~limit, it becomes
${\lim_{T\to 0}\free_H(V,T)=\E_0(V)+\sum_{\bk\mu}\frac{1}{2}\hbar\omega_{\bk\mu}}$.
However, if phonons are treated classically, as in our MD simulations, 
the zero-point energy term does not exist and we
simply have ${\lim_{T\to 0}\free_H(V,T)=\E_0(V)}$.
Therefore the \templim~limits of the mode frequencies calculated within
the QHA and the QPA are different.
They are the harmonic frequencies calculated at
the volumes that minimise ${\E_0(V) + \sum_{\mathbf{k} \mu} \frac{1}{2}\hbar\omega_{\mathbf{k} \mu}(V)}$
and ${\E_0(V)}$, respectively.

\subsubsection{Quasiharmonic approximation (QHA)} 
\label{section:qha}
One of the most important ways in which the vibrational spectra of materials 
change with $T$ is via thermal expansion: Lengthened bonds tend to be
weakened bonds, and weakened bonds vibrate with lower frequencies. Therefore
thermal expansion tends to lower phonon frequencies, on average.
Within perturbation theory, this effect is often modelled using the QHA, which
entails calculating the normal mode frequencies at a range of volumes and 
using them in Eq.~\ref{eqF} to calculate ${\free_H(V,T)}$.
The volume ${V_{\text{min}}(T)}$ that minimizes ${\free_H(V,T)}$ 
is then found and treated as the equilibrium volume at temperature $T$, 
and the normal mode frequencies at this volume are used  to calculate
thermodynamic properties 
from $\free_H(V_{\text{min}},T)$. In the most commonly used form of the QHA~\cite{Karki2000, Calandrini2021},
which is the form that we use, the dependence of ${V_{\text{min}}}$ on $T$ is ignored
when taking derivatives of $\free_H(V_{\text{min}},T)$ with respect to $T$.

The QHA approximates the shift of phonon frequencies by thermal expansion, which is often the largest 
effect of anharmonicity on vibrational spectra at low $T$. However it does not explicitly describe
any phonon-phonon interactions and therefore cannot describe other important effects of 
anharmonicity, such as the broadening of peaks in the mode spectra, ${f_{\bk\mu}}$, 
as $T$ increases.

\subsubsection{Quasiparticle approximation (QPA)}
\label{section:qpa}
At very high $T$, such as in the liquid, each mode spectrum $f_{\bk\mu}$
is not strongly peaked at any particular frequency. However,
in the \templim limit, it becomes the delta function
${f_{\bk\mu}(\omega)=\delta(\omega-\omega_{\bk\mu})}$
and at small finite values of $T$ it is a sharply-peaked
distribution of finite width.

We denote the peak position of mode distribution
${f_{\bk\mu}}$ at temperature $T$ and volume $V$ by
\begin{align}
\bomega_{\bk\mu}(V,T)\equiv \omega_{\bk\mu}(V) + \Delta\omega_{\bk\mu}(T),
\label{eqn:bomega}
\end{align}
where ${\omega_{\bk\mu}(V)}$ is the harmonic 
frequency in the ${T\to 0}$ limit at volume $V$ and ${\Delta\omega_{\bk\mu}(T)}$
is a finite-$T$ correction, which does not
depend on ${V}$ to leading order in anharmonicity~\cite{wallace}.
At constant pressure, $V$ is determined by $T$; therefore
we can write ${\bomega_{\bk\mu}(V,T)=\bomega_{\bk\mu}(V(T),T)=\bomega_{\bk\mu}(T)}$.
The shifted frequencies, ${\bomega_{\bk\mu}(T)}$, are the QP frequencies. They can
be used to calculate thermodynamic properties in manner that is similar, 
but not identical for all properties~\cite{wallace}, 
to how harmonic phonon frequencies are used.

\subsection{Band broadening}
As discussed, 
${f_{\bk\mu}}$ is a delta function 
in the \templim~limit and, as 
$T$ increases from this limit, its first effects on ${f_{\bk\mu}(\omega)}$ are 
to broaden it and to shift it in frequency.

If phonons were independent entities that are 
created and annihilated in sudden random occurrences, which 
might be described as {\em collisions} or {\em scattering events}, 
then, at thermal equilibrium, the average rate at which phonons of each mode ${\bk\mu}$
were created would be equal to the rate at which they were annihilated; and 
annihilation events would be Poisson distributed. It follows that phonon lifetimes
would be exponentially distributed and that each mode correlation function
${C_{\bk\mu}(\bR,t)}$ would decay exponentially as a function of time~\cite{riley_hobson_bence_2002}.

In the \templim~limit, ${C_{\bk\mu}(\bR,t)}$ is sinusoidal
in both $\bR$ and $t$. Therefore, when annihilation events
are Poisson distributed at finite $T$, it becomes an
exponentially-decaying sinusoid, ${C_{\bk\mu}(\bR,t) \sim e^{-\gamma_{\bk\mu} t}\cos \left(\omega_{\bk\mu} t\right)}$, and its Fourier transform, 
${\hE_{\mu}(\bk,\omega)}$, is Lorentzian. 
Therefore ${f_{\bk\mu}(\omega)}$ is also Lorentzian, i.e., 
\begin{align}
f_{\bk\mu}(\omega) \approx \frac{A_{\bk\mu}}{(\omega-\bomega_{\bk\mu})^2 + \gamma_{\bk\mu}^2},
\label{eqn:lorentzian}
\end{align}
where ${A_{\bk\mu}}$ is a constant, 
${\bomega_{\bk\mu}}$ is the QP frequency, 
${2\gamma_{\bk\mu}}$ is
the full-width at half maximum (FWHM) of the Lorentzian
and 
${\gamma_{\bk\mu}}$ is the rate of exponential decay of the 
energy of a phonon QP of mode ${\bk\mu}$. 
Therefore, when ${f_{\bk\mu}}$ can be fit closely by a Lorentzian,
${\gamma_{\bk\mu}}$ quantifies the degree to which it
has been broadened by temperature.

On the other hand, when phonons are treated as lattice
waves that continuously exchange energy, 
each $T$-broadened spectrum, ${f_{\bk\mu}(\omega)}$, is only Lorentzian
at very low $T$ when it is a very narrow peak.
At higher $T$, its shape is determined by the relative
strengths of its couplings to other modes. 

At any finite $T$, if ${\Delta\omega>0}$ is sufficiently small
we can interpret ${f_{\bk\mu}(\omega)\Delta{\omega}}$
as the probability that the frequency of the
oscillation along the eigenvector of mode ${\bk\mu}$
at a randomly chosen time is between ${\omega-\frac{1}{2}\Delta\omega}$
and ${\omega+\frac{1}{2}\Delta\omega}$.
We could also interpret it as the probability that
the duration of the complete oscillation that begins
at a randomly chosen time is between 
${2\pi/(\omega+\frac{1}{2}\Delta\omega)}$
and
${2\pi/(\omega-\frac{1}{2}\Delta\omega)}$.
Therefore we can use Shannon's theorem to 
quantify the effects of $T$ on mode $\bk\mu$
by quantifying the degree of uncertainty in its frequency.
As Shannon demonstrated in the context of signal processing~\cite{shannon}, 
the correct way to quantify this uncertainty is by
the {\em mode entropy}, 
\begin{align}
S_{\bk\mu} \equiv -\sum_{\omega} f_{\bk\mu}(\omega)\log f_{\bk\mu}(\omega).
\label{eqn:entropy}
\end{align}
For a given variance of ${f_{\bk\mu}}$, the shape
that maximises ${S_{\bk\mu}}$ is a Gaussian. 

Although using the mode entropy, ${S_{\bk\mu}}$, to quantify the degree of
broadening is both more general,
and better justified theoretically, than fitting to a Lorentzian, the latter is more common in
the experimental literature.
Therefore we calculated both ${S_{\bk\mu}}$ and ${\gamma_{\bk\mu}}$.

\subsection{Anharmonicity}
\label{section:anharmonicity}
Phonon-phonon interactions can be considered explicitly in perturbation theory 
by including cubic and higher-order terms in the truncated Taylor expansion
of the potential energy.
However, at any chosen order, phonon perturbation theory fails
when $T$ is high enough.
Furthermore, although in the \templim~limit each individual term of order ${Q^m}$ 
contributing to ${\E_m}$ in Eq.~\ref{eqn:energy_expansion} is
larger than each individual term of a higher order ${Q^p}$
contributing to ${\E_p}$, 
there are more nonvanishing ${Q^p}$ terms than ${Q^m}$ terms, in general.
For example, as Ashcroft and Mermin~\cite{Ashcroft_and_Mermin} point out,
the number of quartic terms that do not vanish 
by symmetry can be much larger than the number of nonvanishing cubic terms. As a result, 
even at low $T$, the magnitude of ${\E_4}$ can be comparable to, or greater than, 
the magnitude of ${\E_3}$.
They also point out that a crystal would not be stable if its 
Hamiltonian was ${\E\equiv \E_0+\K+\E_2+\E_3}$; $\E_4$ must be added
for stability.

The fact that ${\abs{\E_m}}$ is not necessarily
greater than ${\abs{\E_p}}$ when ${p>m}$ makes 
truncation of Eq.~\ref{eqn:energy_expansion}, 
which is a necessary step in any application of perturbation
theory to real materials at finite $T$, formally unjustified.
The method based on correlation functions that we use
is free of such complications. Therefore it is a useful tool for 
checking when/whether perturbation theories are applicable
and for assessing their accuracies.

A related limitation of perturbation theory is that
it is often necessary to calculate and analyse so many phonon interaction
terms that the underlying physics can become lost in the clutter.
For example, the anomalous broadening of the longitudinal optical (LO)
mode near the BZ center evident in Fig.~\ref{figr1} has been studied
by perturbation theory and explained, in part, as a consequence
of the band structure~\cite{Giura2019,Calandrini2021}: the number of nonvanishing
contributions to ${\E_3}$ that involve the LO mode near ${\Gamma}$ is very high
because of the locations in \wk-space of the other modes.

This is an important observation, which is likely to be part of 
a complete explanation, but on its own it is not a complete and satisfactory explanation.
Explaining one feature of a spectrum as being a consequence of its other features is not satisfactory
because the spectrum as a whole is not self-determined: It is determined by interactions between waves
passing through a lattice of ions. A more complete explanation of a spectral feature would explain it
in terms of the motions of waves and ions.

It is also important to note that, 
when treating phonon ${\bk\mu}$ at finite $T$ as a QP
whose dynamics are damped with a decay constant ${\gamma_{\bk\mu}}$,
the contribution of modes ${\newbq\nu_1}$ and ${\bbq+\newbbq\,\nu_2}$
to the value of ${\gamma_{\bk\mu}}$ could be negligible despite
the magnitude of the term in ${\E_3}$ proportional
to ${Q_{\bk\mu}Q_{\newbq\nu_1}Q_{\bbq+\newbbq\,\nu_2}}$
being  very large.
It is not enough that modes ${\newbq\nu_1}$
and ${\bbq+\newbbq\,\nu_2}$ exchange a lot of 
energy with mode ${\bq\mu}$: they must do so {\em irreversibly}.
This means that if they absorb some of mode ${\bq\mu}$'s energy, they 
must dissipate it, by coupling to more modes, before  
it has time to return to mode ${\bq\mu}$.
For example, when two
harmonic oscillators are coupled, their combined energy
oscillates back and forth between them.
There are times when one of them has all of the energy
and times when the other has it all.
Therefore, in a thermal population of phonons, it
is not necessarily possible to calculate
decay constants accurately
as sums of few-phonon contributions.

Reference~\cite{Tangney2006} provides an illustration of this point.
It was found that when two coaxial nanotubes are in relative sliding
motion, phonons are resonantly
excited at {\em every} sliding velocity. However, 
it is only at a small number of velocities that these
resonances manifest as a strong friction force
that slow the sliding motion down. The energy exchange
between the mechanical motion of the tubes
and most of the resonantly-excited phonons 
is equitable. High friction only occurs at velocities
for which the resonantly-excited phonons dissipate
the energy they absorb from the sliding motion
more quickly than they return it to that motion.

{\em Irreversible} dissipation of energy always requires
the participation of very large numbers of modes. 
Therefore because, in practice, perturbation theories
are limited to considering few-phonon processes, 
they have a fundamental limitation that
the MD-based method used here and in Ref.~\cite{Tangney2006}
does not share.

The purpose of this section is  not to denigrate
phonon perturbation theories. It is to point out that
both perturbation theories and the 
correlation function approach to calculating spectra
are limited, but in different ways;
therefore they complement one another. We have
used harmonic phonon theory extensively to analyse
our spectra, and it is likely that we would have learned 
much more with the help of anharmonic terms.

In Sec.~\ref{section:acoustic_disruption} we provide
a simple explanation of the anomalous LO broadening, in terms of ions and waves, and without
referring to any other feature specific to the vibrational spectrum of MgO and the other similar
materials in which anomalous LO broadening has also been observed~\cite{Woods1960, Woods1963, Raunio_1969, Raunio_1969_2, Lowndes_1970, Chang_1972, Nilsson_1981, Cowley_1983, Shen_2020, Shen_2021}.

We suggest that the {\em acoustic warping} mechanism we propose in Sec.~\ref{section:acoustic_disruption}
modulates the frequencies of LO modes, and therefore the expectation values of their 
occupation numbers. At thermal equilibrium, a mode whose frequency is fixed exchanges energy with other
modes equitably:  its {\em net} rate of energy exchange with other modes vanishes.
However, because the LO mode's frequency $\loangfreq$ is changing, the expectation value,
${\left[\exp\left(\hbar\loangfreq(t)/k_B T\right)-1\right]^{-1}}$, 
of its occupation number is changing, and this might make its net rate of energy exchange 
with other modes finite. This net exchange of energy may happen quickly 
(e.g., for the reasons explained by Giura {\em et al.}~\cite{Giura2019}) 
relative to the
period of the acoustic mode modulating $\loangfreq$, and so it may
contribute significantly to the degradation of lower-frequency bands.

\section{Simulation details}
\label{section:methods}
\subsection{Atomistic force field}
Atomistic molecular dynamics (MD) and lattice dynamics (LD) simulations were performed 
using a polarizable-ion potential of the form 
described in Refs.~\cite{Tangney_Scandolo_2002_2, Tangney_TiO2_2010, Sarsam_Finnis_Tangney_2013}, 
and used to study the pressure dependence of the melting temperature of MgO in Ref.~\cite{Tangney_Scandolo_2009}.

Our method of force field construction~\cite{Tangney_Scandolo_2003} is a form 
of supervised machine learning~\cite{machine_learning_2020,machine_learning_2021}, albeit one
that predates the widespread adoption of the term {\em machine-learning} in this context.
However, the interactions described by most machine learning force fields are {\em near-sighted}~\onlinecite{machine_learning_2020}, 
whereas a realistic description of long-ranged Coulomb interactions is essential when studying zone
center LO phonons. Therefore, we did not impose any accuracy-lowering near-sightedness constraint
on the mathematical form of our potential.

Our force field's parameters were fit to an effectively-infinite dataset of forces, energy differences and 
stress tensors from density functional theory (DFT) calculations using the PBEsol functional~\cite{PBEsol}. 
As discussed in Refs.~\cite{Madden_Wilson_1996, Madden_Wilson_2000} and~\cite{Tangney_Scandolo_2003}, describing the polarizability
of oxide anions, either implicitly or explicitly, is necessary for an atomistic model of an oxide to accurately describe
the long range fields that are intrinsic to LO phonons.
However, cations' electrons tend to be much more tightly bound and we did not find a significant improvement in the fit to DFT data when 
Mg cations were polarizable, so we assigned a polarizability of zero to them.

The mathematical form of the potential and the values of the parameters used are quoted in Appendix~\ref{MDFF}.
In brief, it is the sum of a pairwise interaction, comprising a Morse potential and a ${1/r}$ Coulomb 
interaction, and the Coulomb interactions between dipoles induced on oxygen anions
and the charges and induced dipoles of other ions. 
The dipole moment of the $i^\text{th}$ oxygen anion 
is expressed as ${\mathbf{p}_i=\mathbf{p}_i^{\text{SR}}+\mathbf{p}_i^{\text{LR}}}$, 
where ${\mathbf{p}^{\text{SR}}_i}$ is a short-range (SR) contribution 
caused by asymmetry of the space in which its 
electron cloud is confined by its six cation neighbours,  and 
${\mathbf{p}_i^{\text{LR}}}$ is the dipole moment induced
by the local electric field (${\mathbf{E}_i}$) from the charges
and dipole moments of all other ions.
At each step of the MD simulation, we first use the method of Wilson and Madden to calculate
${\mathbf{p}^{\text{SR}}_i}$~\cite{Wilson_Madden_1993} as a function
of the distances of ion $i$ to neighbouring ions.
Then we iterate the coupled equations (one for each anion) 
${\mathbf{p}_i = \mathbf{p}^{\text{SR}}_i + \alpha\mathbf{E}_i[\{\mathbf{p}_j\}_{j\neq i}]}$
to self consistency in the set of all dipole moments ${\{\mathbf{p}_i\}}$, where
oxygen's polarizability, $\alpha$, is among the parameters fit to DFT data. Finally, we calculate
the Coulomb energy of interaction between each dipole moment and the charges and dipole moments of all other
ions.

\subsection{Molecular dynamics}
\label{section:MD}
We simulate under periodic boundary conditions and 
our supercell is a ${10 \times 10 \times 10}$ repetition of the two-atom rhombohedral primitive unit cell.
We use velocity rescaling, followed by ${\approx 10\,\mathrm{ps}}$ of equilibration in the ${NVE}$ ensemble, to 
prepare for production runs at each temperature, $T$. 
Our production runs of ${\tmax\approx 100\,\mathrm{ps}}$ are also performed in the ${NVE}$ ensemble 
and the reported values of $T$ are calculated from the average kinetic energy of the production run.
We use a time step of ${0.725\,\mathrm{fs}}$ and sample positions and velocities every ten steps for later analysis.
We chose to perform our MD simulations at constant volume, but with ${P\approx 0}$, instead
of at a fixed average pressure and a variable volume.
Although the magnitudes of the fluctuations of each primitive
cell's volume would be reduced, to some degree, by simulating with the supercell volume fixed, 
this was deemed preferable to polluting our spectra with unphysical artefacts of an MD barostat.

We performed one set of simulations with the volume ${V(T)}$ at each $T$ chosen such 
that the average pressure ($P$) in the crystal at that $T$ is close to zero. The averages of ${(P,T)}$
in these simulations were
${(100\;\text{K}, 0.01\;\text{GPa})}$, 
${(301\;\text{K},-0.01\;\text{GPa})}$, 
${(502\;\text{K},-0.02\;\text{GPa})}$, 
${(1004\;\text{K}, 0.16\;\text{GPa})}$, 
${(1995\;\text{K}, 0.10\;\text{GPa})}$, 
${(2998\;\text{K},-0.06\;\text{GPa})}$, 
${(3521\;\text{K}, 0.16\;\text{GPa})}$, 
and
${(3813\;\text{K}, 6.7\;\text{GPa})}$.
The pressure is large in the highest temperature simulation because the crystal has melted and we 
did not repeat the simulation with the volume adjusted for the liquid phase.
At all lower temperatures the simulation cell is crystalline and the maximum estimated
percentage error in the simulated volume is ${|\Delta V/V| =  \Delta P / B \approx 0.1 \%}$, 
where ${B\approx 160~\text{GPa}}$ is its bulk modulus under ambient conditions~\cite{Karki2000, Fei_1999}.

Another set of MD simulations at approximately the same values of $T$
was performed with the volume fixed at its value in the low temperature
limit, i.e., at the value obtained by minimizing the enthalpy with respect to
atomic positions and the lattice parameter.
The averages of ${(P,T)}$ in this set of simulations were
${(100\;\text{K}, 1.1\;\text{GPa})}$, 
${(301\;\text{K}, 2.4\;\text{GPa})}$, 
${(506\;\text{K}, 3.8\;\text{GPa})}$, 
${(1008\;\text{K}, 6.9\;\text{GPa})}$, 
${(1982\;\text{K}, 12.8\;\text{GPa})}$, 
${(2986\;\text{K},18.6\;\text{GPa})}$, 
and \\
${(3458\;\text{K}, 21.3\;\text{GPa})}$.

\subsubsection{Correlation functions and their Fourier transforms}
We used {\em fast Fourier transforms} (FFTs) to compute vibrational spectra from correlation functions.
We calculated spatial correlations up to the maximum distance possible with our 
supercell, which is ${L=5a}$, where $a$ is the primitive lattice parameter. 
We used the entire production run trajectory of length $\tmax$ to calculate temporal correlations.
The resolutions, ${2\pi/L}$ and ${1/\tmax}$,
with which spectra can be calculated as functions of wavevector ($\mathbf{k}$) and frequency ($f$),  respectively, 
are determined by the sizes of the domains in space ($L$) and time ($\tmax$), respectively, on which the correlation functions are
calculated.
With our supercell we are able to sample five commensurate $k$-points 
between $\Gamma$ ($[0,0,0]$) and $X$ ($[0.5,0,0.5]$), where 
${\mathbf{k}}$ is a
{\em commensurate} $k$-point if waves with wavevector $\mathbf{k}$
respect the periodic boundary conditions, i.e., if 
${\mathbf{k}=m_1\bk_1+m_2\bk_2+m_3\bk_3}$, where each $m_i$ is an integer, 
$\bk_i$ is parallel to the $i^\text{th}$ supercell
lattice vector, and if the length of that lattice vector is an
integer multiple of ${\lambda=2\pi/\abs{\mathbf{k}_i}}$.
The frequency resolution of our raw spectra is ${0.01\;\text{THz}}$, but we smooth
them along the frequency axis
by convolving them with Gaussians of standard deviation
${0.05\;\text{THz}}$, for the spectra at ${T\lesssim 1000\;\text{K}}$, and
${0.1\;\text{THz}}$ for the spectra at ${T\gtrsim 2000\;\text{K}}$.

\subsection{Lattice dynamics}
To calculate the phonon band structure within the harmonic approximation, we find
the equilibrium (${T\to 0}$) structure of the supercell and then calculate
the dynamical matrix from finite differences of forces after displacing
atoms slightly from equilibrium. To avoid artefacts of interpolation, 
we only plot the frequencies
at commensurate $k$-points, but we increase their density in reciprocal
space by calculating phonon spectra on
${M\times M\times M}$ supercells for ${M\in\{6,7,8,9,10,11\}}$.

As discussed in Sec.~\ref{section:qha}, the QHA simply amounts to 
calculating the phonon band structure at the equilibrium volume 
for the given temperature, $T'$, rather than at the equilibrium
volume of the \templim limit.
The equilibrium volume at temperature $T'$ is the volume
that minimizes the free energy; but we cannot calculate the free energy easily, precisely, 
or accurately due to important contributions to it from improbable microstates.
Improbable microstates are unlikely to be sampled in an MD simulation
and if, by chance, they did occur, they would be oversampled.
However we can calculate the free energy under the simplistic assumptions
that phonons do not interact with one another and that the vibrational energy is distributed
among the phonon bands according to the Bose-Einstein distribution. Therefore, when
we apply the QHA we assume that band theory is a good approximation and we calculate
the phonon bands as a function of volume. We use these frequencies
in Eq.~\ref{eqF}, which is exact in the \templim limit where band theory
is exact, to find the volume that minimizes ${F(V,T) \eval_{T=T'}}$.

We calculated ${F(V,T)\eval_{T=T'}}$ for a range of volumes corresponding
to a set of lattice parameters with uniform spacing ${\SI{0.02}{\angstrom}}$,
and we interpolated between these values with cubic splines.
To calculate the phonon frequencies at each volume we use a ${10\times 10\times 10}$ supercell 
to calculate the dynamical matrix 
at each $\bk$ in the ${10\times 10\times 10}$ commensurate set
from finite differences of forces.
When calculating ${F(V,T) \eval_{T=T'}}$ from Eq.~\ref{eqF}, 
the sum over $\bk$ is a sum over the set of commensurate $\bk$-points.

\section{Results} 
\label{section:results}
\begin{table*}[!]
    \centering
    \begin{tabular}{c|c|c|c|c|c|c|c|c}
        & \SI{0}{\kelvin}  & \SI{100}{\kelvin}  &  \SI{300}{\kelvin} & 500 K & 1000 K & 2000 K & 3000 K & 3500 K  \\
       \hline
       ${a_{\text{QHA}}/\angstrom}$ & $4.2294 \;\;(0.0)$ & $4.2296\;\;(0.02)$ & $4.2350\;\;(0.40)$ & $4.2442\;\;(1.05)$ & $4.2727\;\;(3.10)$ & $4.3558\;\;(9.24)$ & - & -  \\
       ${a_{\text{MD}}/\angstrom}$ & $4.2119 \;\;(0.0)$ & $4.2202\;\;(0.59)$ & $4.2305\;\;(1.33)$ & $4.2410\;\;(2.09)$ & $4.2670\;\;(3.98)$ & $4.3296\;\;(8.62)$ & $4.4120\;\;(14.9)$ & $4.4640\;\;(19.1)$ 
    \end{tabular}
    \caption{Comparison between the average lattice constants in our
  constant pressure MD simulations, ${a_{\text{MD}}}$, and the {\em quasi}harmonic lattice constants, ${a_{\text{QHA}}}$,  
  at the same temperatures. The numbers in brackets are the percentage changes in volume with respect
to the ${T\to 0}$ volume.}
    \label{tabm1}
\end{table*}
We presented the central result of this work in Fig.~\ref{figr1}, which contains
eight kinetic energy spectra that show how thermal disorder degrades the optical bands of MgO.
Before analysing and discussing these spectra, we discuss some
limitations of our calculations that should be borne in mind while interpreting them.

\subsection{Sources of inaccuracy and imprecision}
\label{section:limitations}
The first limitation is our use of approximate interatomic forces. 
The accuracy with which  our force field calculates phonon frequencies 
at ${\SI{300}{\kelvin}}$ is evident in the bottom-right panel of Fig.~\ref{figr1}, where
we compare with measured phonon frequencies and with those calculated {\em ab initio} with DFPT.
The underestimation of LO frequencies at small wavevectors is common for
force fields of this kind~\cite{Liang_2006, Tangney_Scandolo_2003, Sarsam_Finnis_Tangney_2013},
and has been discussed in Ref.~\onlinecite{Sarsam_Finnis_Tangney_2013}. It is likely to be caused 
by the induced dipoles of the force field {\em over}screening the long wavelength electric fields that 
LO phonons create, and which increase the frequencies of LO phonons relative
to transverse optical (TO) phonons. 

The accuracy with which our force field describes thermal expansion is evident
in Table~\ref{tabm1}. 
Its accuracy for other properties and at higher $T$ is more difficult
to assess because we have neither more accurate calculations nor experimental measurements
to compare with. However, force fields of the same
mathematical form, and parameterized in the same way, were used
in Refs.~\onlinecite{Tangney_Scandolo_2003} and~\onlinecite{Tangney_Scandolo_2009} and shown 
to predict the pressure dependences of the crystal's volume ($V$) and melting temperature ($T_m$) 
accurately, and to produce pair correlation functions for molten MgO 
in perfect agreement with those produced by {\em ab initio} MD simulations.

To parameterize the force field used in this work we fit to DFT data calculated with the PBEsol functional~\cite{PBEsol}, which 
tends to be more accurate than the local density approximation used in
Refs.~\onlinecite{Tangney_Scandolo_2003} and~\onlinecite{Tangney_Scandolo_2009}.  Furthermore, 
we fit the parameters to DFT calculations of {\em crystalline} MgO, whereas
those used in Refs.~\onlinecite{Tangney_Scandolo_2003} and~\onlinecite{Tangney_Scandolo_2009} were required
to describe MgO in both molten and crystalline forms, which meant compromising
on the accuracy with which each phase was described.

Based on the tests performed here and in those previous works we suggest
that, at worst, our force field should be regarded as describing
an MgO-like material. We further suggest that some of the gross features 
of the $T$-dependence of the spectra calculated with it, including those
that we have selected for discussion below, are caused by simple physical
mechanisms that would occur in other ionic crystals and therefore 
might manifest in their vibrational spectra.


A second limitation of our calculations is that our MD simulations
are classical simulations. Therefore their accuracy at low
$T$ is questionable. However, when $T$ is
comparable to, or greater than, the Debye temperature (${T_D\sim \SI{1000}{\kelvin}}$), 
which is the range of most interest for studying the breakdown of band theory, 
this approximation should not affect our results significantly. Furthermore, 
we will show in Sec.~\ref{section:thermal_expansion} that the heat capacities calculated
from the  quasiparticle energies extracted from MD simulations are
in very good agreement with measurements at all values of $T$ between
${\SI{300}{\kelvin}}$ and ${\SI{1800}{\kelvin}}$.

We regard our use of a supercell of finite size in our MD simulations
as by far the most important source of 
inaccuracy and imprecision in the kinetic energy spectra presented.
It means that the only lattice waves present our 
MD simulations were those whose wavevectors are commensurate with the supercell.
Our use of a ${10\times10\times10}$ supercell means that
the longest finite wavelength among the phonons in our simulation was only ten times 
the length of a primitive lattice vector.
To make clear that we only calculate spectra at a finite number of points
in \wk-space, we have pixelated the spectra, with one
pixel centered at each commensurate $\bk$-point.
However, the absence of any vibrations at incommensurate \wk-points
would change how energy is distributed among the commensurate
set of points. For example, in a real crystal there would be far more channels (modes)
through which energy and momenta could be exchanged.

\subsection{Deviations of finite-$T$ spectra from band structures}
\label{section:deviations}
Let us now begin discussing Fig.~\ref{figr1}, which compares the full
\wk-resolved kinetic energy spectra, ${\hE^\K(\bk,\omega)}$, from our MD simulations at different values of $T$,
after each one has been normalized so that it integrates to one. 

As mentioned in Sec.~\ref{section:theory_spectra}, the equipartition theorem implies 
that if the spectra were converged fully with
respect to simulation time, this normalization would be equivalent to dividing
each one by the same constant and by $T$.
Regions of high and low energy density are coloured dark blue and white, respectively, with
the same colour scale used at each $T$.
For comparison, the \templim band structure and the finite-$T$ QHA band structures
are plotted over the full spectrum with red and green triangles, respectively.

The spectra show a progressive transition between two limits of $T$:
Phonon bands are well defined at low $T$, whereas at very high $T$ they 
are much less well defined or not defined. At ${\SI{300}{\kelvin}}$ the phonon dispersions do not differ substantially from the \templim bands
calculated within the HA: All of the vibrational energy is localized near the normal 
mode points, ${(\bk,\omega_{\bk\mu})}$, and the widths of the peaks at these points are small.
At ${\SI{3800}{\kelvin}}$, on the other hand, the phonon bands have vanished and the spectrum is much more uniform and
has little observable structure. 
This is because the crystal has melted and because spatial and temporal correlations are very  
short in a liquid. Therefore the phonon assumption of atoms moving collectively as waves has 
broken down completely.

The spectra at other $T$s show various stages of the progression 
between the low-$T$ limit, in which 
atoms move as lattice waves with well defined frequencies and wavevectors, and the liquid, in which each atom moves
independently of all other atoms, except those closest to it. 

Note that MgO melts at ${T_m\lesssim \SI{3100}{\kelvin}}$~\cite{Tangney_Scandolo_2009, Kimura_2017}, but it is well known 
that, by imposing a degree of long range order, the periodic boundary conditions used in MD simulations 
can prevent melting until $T$ is significantly larger than  $T_m$~\cite{Belonoshko_1996}.
Therefore, the ${T=\SI{3500}{\kelvin}}$ spectrum in Fig.~\ref{figr1} should be regarded
as the spectrum of a superheated MgO crystal.

\subsubsection{Selected features of the kinetic energy spectra}
There is a lot of complexity in the $T$-dependence of the full spectrum (Fig.~\ref{figr1})
and a much more extensive and detailed study would be required to explain it all. 
Therefore we provide the data used to 
produce Fig.~\ref{figr1} 
as Supplementary Material
so that others may analyze it further~\cite{supplementary}.
We focus our attention on two important gross features of the spectrum's $T$-dependence, and on one
striking specific feature.

The first gross feature is that all phonons
shift to lower frequencies as $T$ increases. Most of this softening can be attributed to the weakening
of bonds by thermal expansion. We will discuss this in more detail
in Sec.~\ref{section:thermal_expansion}. 

The second gross feature is that optical bands, and particularly LO bands, 
lose definition much more rapidly with increasing $T$ than acoustic
bands; and zone center modes (${\bk\to 0}$, $\lambda\to \infty$) lose definition much more rapidly than
zone boundary modes. 
Much of the intensity seen at high frequencies 
and small wavevectors at low $T$ gradually moves
towards lower frequencies and larger wavevectors as $T$ increases.
This is easiest to see along the wavevector paths ${\Gamma\to X}$ and
${\Gamma\to L}$, which are straight line segments connecting the zone 
center ($\Gamma$) to the high symmetry points $X$ and $L$ on the zone boundary.
We will denote the wavevectors at $X$ and $L$ by ${\bk_X}$ and ${\bk_L}$, respectively.

When analysing spectra, we will ignore the point ${\Gamma}$ itself, which does not represent the
limit ${\bk\to 0}$, but the point ${\bk=0}$: 
The intensity at $\Gamma$ in Fig.~\ref{figr1} is the kinetic energy of rigid relative motion
of the Mg and O sublattices in our simulations, which does not interest us.
However, note the disappearance, at high $T$, of most of the intensity at optical frequencies and at
wavevectors of
${\frac{1}{5}\bk_X}$, ${\frac{2}{5}\bk_X}$, ${\frac{1}{5}\bk_L}$, and ${\frac{2}{5}\bk_L}$, which are 
the smallest finite wavevectors along these paths that our simulation supercell can accommodate.
The redistribution of intensity away from high frequencies at these
wavevectors means that, at very high $T$, little of the crystal's kinetic
energy exists as {\em coherent} optical waves whose wavelengths are greater
than about five or ten lattice spacings. 
The TO branches retain 
significant amounts of energy at these wavelengths, 
but the LO bands have almost vanished.
As discussed in Sec.~\ref{section:T_effects}, much of this is a manifestation
of thermal disorder reducing the correlation lengths 
and times of collective motions along the cell eigenvectors of the affected modes, 
causing them to become less wavelike and more localized in spacetime.

While this is happening to optical bands at small wavevectors, 
the acoustic branches of the spectrum remain relatively
robust. Melting manifests in the spectra as the
sudden total loss of definition and integrity of the acoustic bands
between ${T=\SI{3500}{\kelvin}}$ and ${T=\SI{3800}{\kelvin}}$.

The broadening of bands does not {\em necessarily} mean that the waves contributing
to the band are less coherent. A band of finite width can broaden further without
the phonons' coherence lengths and lifetimes reducing further if the broadening is
caused by a very slow modulation of the properties of the underlying lattice.
As we discuss further below, 
the frequency and/or wavevector of an optical mode could be modulated by an acoustic phonon whose
period and wavelength are much larger than the optical phonons' coherence length and lifetime, respectively. 

The specific striking feature of the spectra that we have chosen to discuss
is that
the loss of definition of the optical modes appears to 
begin with an anomalously-large broadening of the LO modes nearest to $\Gamma$, and to 
spread out from these \wk-points as $T$ increases.
The broadening of these modes is even visible at ${\SI{100}{\kelvin}}$, which
is the lowest $T$ at which we calculated the full spectrum, and it occurs for the LO
phonons at all wavevectors $\bk$ near $\Gamma$, regardless of their directions.

We are not the first to observe the anomalous LO linewidth at low $T$ in MgO~\cite{Giura2019, Calandrini2021};
and similar features have
been observed experimentally in similar materials
since the 1960s, and more recently in calculated spectra~\cite{Woods1960, Woods1963, Raunio_1969, Raunio_1969_2, Lowndes_1970, Chang_1972, Nilsson_1981, Cowley_1983, Shen_2020, Shen_2021}.
However, we have not found other studies that show how it evolves as the crystal is heated to a very
high $T$, and that it marks the beginning of the gradual melting of the entire LO band.
When the set of spectra in Fig.~\ref{figr1} are examined collectively, the melting
of the LO band may be the most noticeable feature of the spectrum's $T$-dependence
between ${\SI{100}{\kelvin}}$ and ${\SI{3500}{\kelvin}}$. 

Because the LO band melts gradually and systematically from its apparent origin, at low $T$, 
as a localized anomaly near the BZ center, and because
similar anomalies have been observed in other materials, we believe that
its primary 
cause is a simple and general physical mechanism, which we
refer to as {\em acoustic warping of optical phonon fields}.
The effects of this mechanism on vibrational spectra would be particular pronounced in crystals with large LO-TO splittings, such as
strongly ionic materials with the rocksalt crystal structure.
We explain it in the sections that follow.

We begin by suggesting an explanation for why strong coupling between low frequency acoustic 
phonons and optical phonons would melt optical bands much faster than acoustic bands.
Then we explain the acoustic warping mechanism, why it would cause LO bands to melt, 
and why this melting would begin near the BZ center before gradually
progressing outwards, until the only LO modes that remain are those at the BZ boundary.

\subsubsection{Separation of optical and acoustic timescales in the long wavelength limit}
\label{section:adiabatic_decoupling}
The acoustic branches can be thought of as a skeleton on which the optical branches `hang', 
because
{\em spatially-coherent} countermotion of cations and anions about a reference structure
would not happen unless the reference structure was itself spatially coherent.
For example, consider the projection,
${\braket{\loeig_{\bk}}{\psi_\bR}}$
of the displacement 
of cell $\bR$ onto the cell eigenvector of the LO mode at $\bk$.
When regarded as a function of $\bR$, this projection is very unlikely to have order on length scale ${\lambda}$ if the crystal
does not have order on length scale ${\lambda}$.
Crystalline order is gradually lost as $T$ increases, 
and this manifests as a gradual reduction of the correlation lengths and times
of this projection, i.e., as an increase in the rate of decay of ${\abs{\locorr_\bk(\bR,t)}}$ 
as a function of $\bR$ at fixed $t$ and as a function of $t$ at fixed finite $\bR$.

The energy cost of acoustic distortions protects 
long range order in the {\em time average} of the crystal's structure.
However, if ${\abs{\bk}}$ is small, 
the crystal having long range order when averaged over several
periods of a long-wavelength acoustic (LWA)
mode is not sufficient to protect long range order in ${\braket{\loeig_{\bk}}{\psi_\bR}}$.
It is insufficient because the periods (frequencies) of zone center LO modes
are much shorter (higher) than those of LWA modes.
Therefore long range order of ${\braket{\loeig_{\bk}}{\psi_\bR}}$
as a function of $\bR$
relies, not on the degree to which the 
{\em time average} of LWA mode displacements 
preserve crystalline order (i.e., vanish), but on the degree
to which they preserve it on timescales much shorter than 
LWA mode periods.
Long range order of the optical branches
requires the {\em amplitudes} of LWA
modes to be small, not their time-averaged displacements.

The converse is not true and, in either the limit of large LO frequency, $\loangfreq_\bk$, or the limit of small
LWA frequency, ${\lwaangfreq_\bk}$,
the coupling between LWA and LO modes is almost unidirectional: LO vibrations are highly sensitive
to LWA mode displacements, whereas LWA vibrations are much less sensitive
to LO mode displacements.  
The reason for the high sensitivity of LO phonons to LWA phonons
will be discussed in Sec.~\ref{section:acoustic_disruption}. 

LWA phonons are relatively insensitive to LO phonons because 
an LO mode's period, ${2\pi/\loangfreq_{\bk}}$, is so short that 
LO displacements average to zero in much less than the period, ${2\pi/\lwaangfreq_\bk}$, of an LWA phonon.
Therefore the forces they exert on an LWA mode cancel one another before the LWA mode has
had time to respond to them.
On the length and time scales relevant to the lowest-frequency LWA modes of a macroscopic crystal, the crystal is a continuum
and the LWA modes do not see optical mode disorder directly, but experience its effects indirectly
through the $T$-dependence of the crystal's elastic constants.

This partial decoupling of a slow-moving degree of freedom from a much faster 
one is known as {\em adiabatic decoupling}~\cite{landau, arnold, spohn_2001,Tangney_JCP_2006,gilz_2016}.
It is exploited by the Born-Oppenheimer
approximation:  An excellent approximation to the 
force exerted on a heavy nucleus by electrons
is calculable from the electron density, which can loosely be thought
of as the time average of the electrons' positions.

Adiabatic decoupling does not mean that the LWA modes do not exchange energy with the optical
modes. It means that their energy exchange is so rapid that they barely notice.
The net energy exchange in a time $\tau$ can be very large
if ${\tau < 2\pi/\loangfreq_\bk}$, but is likely to be negligible if ${2\pi/\loangfreq_\bk\ll\tau \ll 2\pi/\lwaangfreq_\bk}$,
because its average over one complete LO period is small and its average over many complete
periods is even smaller.
Optical mode disorder changes so quickly that every complete LWA oscillation occurs in the presence 
of an almost equivalent background of optical displacements, whereas LWA disorder changes so
slowly that every complete LO oscillation occurs in the presence of a unique and inequivalent
background of LWA displacements.

Therefore it is approximately true that
LWA phonons only experience the many rapidly-changing LO displacements as a slight {\em dressing}, 
which changes their frequencies very little.
Nevertheless, and at the same time, if there exists an effective LWA-LO coupling mechanism, 
the frequencies and coherence lengths and times of LO phonons are strongly influenced by whatever
LWA displacements exist during their lifetimes.

Of course, the adiabatic decoupling picture, in which optical modes see `frozen' 
acoustic modes and acoustic modes do not see optical modes because the
time averages of their displacements vanish, 
is very much an idealized limiting case. Adiabatic decoupling of acoustic modes
from optical modes becomes a perfect decoupling in 
the ${\bk\to 0}$ limit, but is likely to be far from perfect at
the smallest finite wavevectors present in our simulations.
Clearly this picture would not apply 
to the transverse acoustic (TA) and TO modes in the
regions of the BZ where they
occupy the same small frequency window between about $\SI{10}{THz}$
and ${\SI{15}{THz}}$. However, it does seem to explain why
the lowest-frequency
acoustic modes are broadened less than the highest-frequency optical modes.

As $T$ increases from the \templim~limit, it is the LWA modes that
are first to become active, because they are lowest in energy.
Therefore, there may be a common explanation for why the LO band 
is the first to melt, or partially melt,
for why zone center modes melt before zone boundary modes, and for why acoustic bands are degraded
less at the highest $T$'s than optical bands:
Acoustic modes are degraded less because of the partial immunity to optical disorder
that adiabatic decoupling affords them. LO modes are not protected
by adiabatic decoupling and,
as we now explain, there is a very simple and effective mechanism by which they can be disrupted,
and their frequencies changed, by acoustic disorder; or even 
by a single quasistatic acoustic perturbation of the crystal.

\subsubsection{Acoustic warping of the LO electric field}
\label{section:acoustic_disruption}
By modulating the relative displacements of cations and anions, 
an LO wave of wavevector ${\bk}$ creates regions
of excess charge at its nodes, from which emanates
an electric field, ${\lofield_\bk}$, of the same wavelength, ${\lolambda_\bk\equiv 2\pi/\abs{\bk}}$.
This field, which we referred to as the LO mode's {\em intrinsic field} above, 
opposes the LO wave's motion, thereby increasing its frequency, ${\loangfreq_\bk}$~\cite{Born_and_Huang, Ashcroft_and_Mermin}.

One way to understand the origin of ${\lofield_\bk}$ is to consider the case 
in which ${\lolambda_\bk}$ is orders
of magnitude larger than a primitive lattice spacing; 
and to imagine partitioning the crystal into primitive
unit cells of dipole moment $\bd$, whose volume we will treat as infinitesimal.

If the LO mode at wavevector
$\bk$ becomes active, it modulates $\bd$ along an axis 
parallel to ${\hbk\equiv \bk/\abs{\bk}}$
by modulating the displacements of 
Mg cations from the O anions with which they share a primitive cell.
On length scale ${\lolambda_\bk\gg \abs{\ba_1}, \abs{\ba_2},\abs{\ba_3}}$,
we can define a local spatial average of the 
cell dipole moment per unit volume, 
${\mathbf{P}\equiv \expval{\bd}/\abs{\ba_1\cdot(\ba_2\times\ba_3)}}$, 
and treat it as a continuous function of position.
The dependence of $\bd$ on the 
choice of primitive cell makes ${\mathbf{P}}$  ill-defined~\cite{resta-vanderbilt-2007};
however its derivatives with respect to space and time are the same for every choice.
Therefore the density ${\brho_b\equiv-\div\mathbf{P}}$
of {\em excess charge} or {\em bound (`b') charge} 
is independent of the choice of primitive cell and is a well-defined
physical quantity.

If only the LO mode at $\bk$ is active, 
${\brho_b}$ varies in direction $\hbk$ with wavelength ${\lolambda_{\bk}}$.
Its magnitude is largest at the wave's {\em nodes}, 
which is where the Mg-O displacement varies most from cell to adjacent cell
along ${\hbk}$. When multiple LO modes are active, ${\brho_b}$ has
a contribution from each one and we will denote the contribution
from the one with wavevector $\bk$ by ${\brhobk}$.

${\lofield_\bk}$ is the field emanating from 
 ${\brhobk}$.
If it was absent or negligible (${\lofield_\bk\approx 0}$) 
the LO and TO modes would have the same frequency in the long wavelength
limit by symmetry~\cite{Born_and_Huang}. Therefore near $\Gamma$, ${\lofield_\bk}$
increases the value of ${\loangfreq_\bk/2\pi}$ by
almost ${\SI{10}{THz}}$. It follows that any weakening
or strengthening of $\lofield_\bk$ could change $\loangfreq_\bk$ quite dramatically:
a ${10\%}$ reduction in $\abs{\lofield_\bk}$ near $\Gamma$ would
reduce $\loangfreq_\bk/2\pi$ by ${10\%}$ of the difference between
the frequencies of the LO and TO modes, which is ${\sim \SI{1}{THz}}$.

A quasistatic acoustic perturbation of the crystal 
changes $\loangfreq_\bk$ by weakening or strengthening $\lofield_\bk$;
and {\em different} quasistatic acoustic perturbations would result in different 
frequencies. Therefore the distribution of the LO mode's energy among frequencies, which is localized
at a single frequency in the \templim limit, should broaden significantly
as $T$ activates the acoustic modes. 

There are some obvious mechanisms by which TA and 
longitudinal acoustic (LA) distortions would change ${\lofield_\bk}$.
The first is simply disorder: Acoustic perturbations whose wavelengths differed from 
${\lolambda_\bk}$ would break the periodicity of ${\lofield_\bk}$ by breaking the
periodicity of ${\brhobk}$. This would reduce the LO wave's coherence and, in many
or most cases, reduce ${\loangfreq_\bk}$ by weakening ${\lofield_\bk}$.

A second mechanism is that a TA or LA perturbation 
with wavevector ${\bk}$ (or ${m\bk}$, where ${m\in\integer}$), 
would change ${\lofield_\bk}$ by perturbing the
charge reservoirs centered at the antinodes of ${\brhobk}$.
A TA perturbation would displace the positive and negative reservoirs
relative to one another along an axis perpendicular to ${\hbk}$, whereas
an LA perturbation would expand or compress them.

In planes perpendicular to $\bk$, the part
of the crystal that a phonon with wavevector ${\bk}$ perturbs is finite in size
at any finite $T$.
If the LO wave's lateral extent was smaller than ${\lolambda_\bk}$, 
a transverse relative displacement of $\brhobk$'s oppositely-charged antinodes 
would change the direction of $\lofield_\bk$ locally. This would reduce
the magnitude of its component along ${\hbk}$, thereby reducing
${\loangfreq_\bk}$. 

An LA perturbation, on the other hand, would 
modulate the magnitude of ${\lofield_\bk}$ along ${\hbk}$.
If, at a given moment, the positive
antinodes of ${\brhobk}$ were compressed (expanded) by an LA wave of wavevector $\bk$, 
its negative antinodes would be expanded (compressed) at that moment.
However the phase velocities of LO and acoustic waves are different, in general, 
which means that the antinodes of an acoustic wave would be moving 
relative to those of ${\brhobk}$, making it a time-dependent perturbation of the LO wave.

A third mechanism is that a quasistatic acoustic perturbation of the crystal would
create regions in which the crystal is compressed and regions in which it is expanded. 
Compressing an ionic crystal increases its Madelung energy and the magnitudes of local
electric fields. 
In MgO this results in the electronic band gap widening under pressure~\cite{MgO_bandgap_pressure}:
It increases the electrostatic potential at oxygen sites, thereby lowering 
the energy of the highest-energy valence electrons.
It has also been shown that ${\loangfreq_\bk}$ increases under pressure~\cite{Ghose_2006}, 
which may be a result of all fields increasing in magnitude when interionic distances are shortened.
Therefore an acoustic wave with a wavelength much larger than ${\lolambda_\bk}$ might
increase $\loangfreq_\bk$ in compressed regions of the crystal and reduce it in expanded regions.

The acoustic-LO coupling mechanisms discussed above suggest that, as soon as
acoustic modes become active, we should see a range of LO frequencies corresponding
to LO vibrations occurring in the presence of different 
backgrounds of acoustic deformations of the crystal.
Let us now examine the spectra presented 
in Fig.~\ref{figr1} for consistency with this prediction.
We will focus on the lowest-$T$ spectra and on 
what happens near $\Gamma$, which is the region of the BZ occupied by
the first modes to become thermally-active as the crystal is
heated from the \templim~limit.
To see clearly how the degradation of the LO band depends on wavevector magnitude, let
us focus again on the paths ${\Gamma\to X}$ and ${\Gamma\to L}$ at the left
hand side and at the right hand side, respectively, of each spectrum in Fig.~\ref{figr1}. 

As $T$ increases, the earliest deviations from a perfect band structure
occur close to $\Gamma$ along both paths (and also along other paths, which we do not discuss).
For example, at ${\SI{300}{\kelvin}}$ and ${\SI{500}{\kelvin}}$
the LO modes at ${\frac{1}{5}\bk_X}$ and ${\frac{1}{5}\bk_L}$ 
have broadened noticeably in frequency.
Along ${\Gamma\to X}$ the LO broadening is accompanied by a noticeable broadening of the LA modes, 
whereas along ${\Gamma \to L}$ it is accompanied by a noticeable broadening of the TA modes.
In the latter case the broadening of the TA mode is smaller 
at ${\bk=\frac{1}{5}\bk_L}$
than it is further away from $\Gamma$, despite the largest broadening of
the LO mode being at ${\bk=\frac{1}{5}\bk_L}$.
However, ${\frac{1}{5}\bk_L}$ is the wavevector at which the difference 
between LO and TA frequencies is greatest, which means that it
is the wavevector at which the adiabatic decoupling of acoustic modes
from optical modes discussed in Sec.~\ref{section:adiabatic_decoupling} would be most effective.
Therefore, the low-$T$ broadening of the LO modes near $\Gamma$,
its progression to larger wavevectors
as $T$ increases, and the broadening of acoustic modes that accompanies it, 
all appear to be consistent with acoustic modes changing
${\loangfreq_\bk}$ and making LO waves less coherent by warping ${\lofield_\bk}$.

The main point of this section (\ref{section:acoustic_disruption})
is that there exists a mechanism for strong coupling between the acoustic and LO modes
of an ionic crystal, which would degrade its LO bands but degrade its acoustic bands to
a much lesser degree. We have provided simple 
physical reasons why we {\em should} see anomalous broadening of the LO mode in our
spectra and, as we analyze the spectra of individual modes
in more detail in Sec.~\ref{section:broadening}, we will provide further indirect evidence that the mechanism we
proposed is responsible for the observed broadening.  
We end this section by summarizing our physical reasoning.
 
It is well known~\cite{Born_and_Huang, wallace, Ashcroft_and_Mermin} that the frequency of a long wavelength LO phonon
is increased by its intrinsic electric field, $\lofield_\bk$.
It is also known that, as $T$ increases from the \templim~limit,
the earliest contributions to thermal disorder come from the lowest energy
modes, which are LWA modes.
It is obvious that, by breaking the long range order of primitive
cell displacements along LO cell eigenvectors, 
acoustic perturbations of the crystal would warp $\lofield_\bk$. It follows
that acoustic phonons would change the frequencies of LO waves and, by disordering
the lattice, make them less coherent. 

As $T$ increases, the first manifestation of this effect
in the spectra would be the LO band near the BZ center beginning to broaden and melt.
As $T$ continues to increase, causing significant activity among higher-energy and larger-wavevector 
acoustic modes, we should expect to see 
the melting of the LO band spread from BZ's center towards its boundary.
All of this is consistent with what is observed in Fig.~\ref{figr1}.

\subsection{Effects of thermal expansion}
\label{section:thermal_expansion}
\begin{center}
\begin{figure*}[!]
    \includegraphics[width=\textwidth]{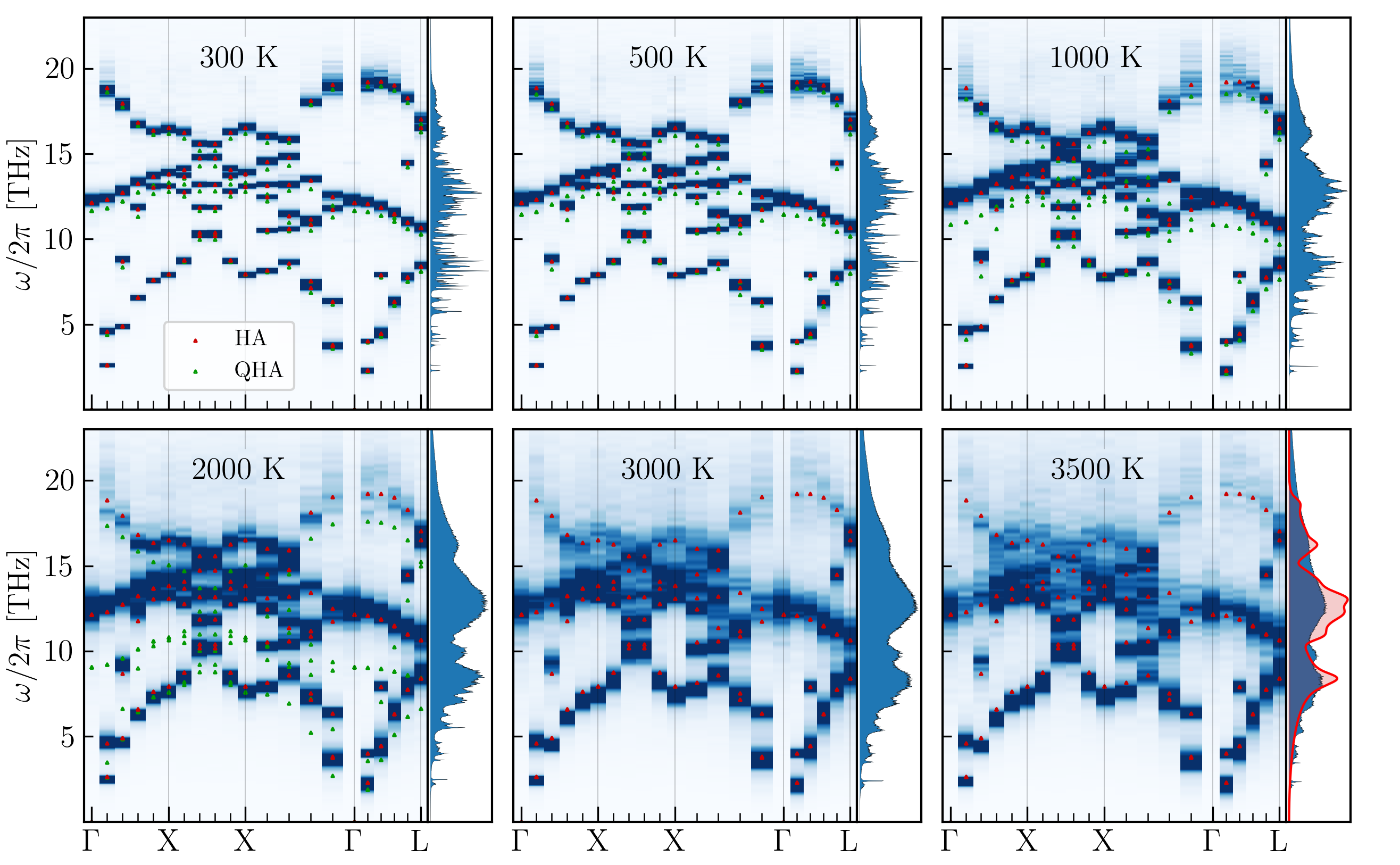}
    \caption{Distribution of vibrational energy in reciprocal 
spacetime when thermal expansion is suppressed by performing simulations at the \templim~volume. The normalization 
of the data and the
colour scale of each plot are the same as those used for each plot in Fig.~\ref{figr1}.
As in Fig.~\ref{figr1}, the red and green triangles are phonon frequencies at zero
temperature and at ${300\;K}$ within the QHA, respectively.
The temperatures are approximate.
}
    \label{fig:fig2}
\end{figure*}
\end{center}
To disentangle the effects of thermal expansion from other mechanisms
by which $T$ changes the spectra, 
let us turn our attention to Fig.~\ref{fig:fig2}, which presents 
the spectra calculated from MD simulations in which 
the volume of the crystal at all values of $T$ was 
constrained at the value ${V(0)\equiv \lim_{T\to 0}V(T)}$.
We will refer to the MD simulations whose spectra are presented in Figs.~\ref{figr1}
and~\ref{fig:fig2} as the ${V(T)}$ and ${V(0)}$ simulations, respectively.

The comparison between Figs.~\ref{figr1} and~\ref{fig:fig2}
demonstrates that the overall softening of most modes 
that occurs when the crystal is free to thermally expand 
does not occur when it is prevented from expanding.
This, and the comparison with the QHA results below, confirm that
thermal expansion bears most of the responsibilty for this softening.

The other features of the spectra that we have chosen to discuss,
namely, that melting of the band structure begins
with the LO mode near $\Gamma$, before spreading to
the BZ boundary and to other modes, 
are present in both Fig.~\ref{figr1} and Fig.~\ref{fig:fig2}. Therefore
we can rule thermal expansion out as their cause.
However it is worth noting that, at very high $T$, the TO branches look much
more like bands in the ${V(0)}$ spectra than they do in the ${V(T)}$ spectra.
This may be because thermal expansion causes the TO and TA branches of
the spectrum to overlap more in frequency, which would strengthen
TO-TA interactions.

The QHA provides a reasonably accurate approximation to the phonon band structure up to at least 
${\SI{500}{\kelvin}}$, albeit one that is not noticeably
closer to the MD band structure than the HA. 
However, its overestimation of the softening of 
TO frequencies
is quite dramatic at ${T=\SI{1000}{\kelvin}}$ and ${T=\SI{2000}{\kelvin}}$.
Referring to Table~\ref{tabm1}, we see that the QHA lattice parameter
is larger than the MD lattice parameter at each $T$. However, 
these differences are much too small to explain the differences in frequency between
the QHA bands and the ${V(T)}$ MD bands. When the $T$-dependent
frequency of each mode is plotted as a function of the ${T}$-dependent volume for
both sets of calculations, the QHA curves lie below
the ${V(T)}$ MD curves and the gap between them widens as $T$ increases.
Thermal expansion must explain the softening of the QHA bands with increasing $T$, because 
the difference in $V$ is the only difference between
calculations performed at different $T$. However the degree to which
the QHA oversoftens bands cannot be explained fully by inaccuracies
of QHA thermal expansivities.
We will return to this issue in Sec.~\ref{section:broadening}.

Karki {\em et al.}~\cite{Karki2000} found
that QHA calculations of several thermodynamic properties started to deviate
significantly from experimental measurements at around ${\SI{1000}{\kelvin}}$, 
with both thermal expansivity and the isobaric heat capacity ${C^{(P)}}$ being 
overestimated at high $T$ and the degrees to which they were overestimated
increasing with $T$. 
Therefore, to investigate whether or not this is a failure of 
the quasi-independent phonon assumption 
we have calculated the heat capacity within the QPA~\cite{wallace}
For each mode ${\bk\mu}$ we fitted the mode spectrum, 
${f_{\bk\mu}(\omega)}$, with a Lorentzian (Eq.~\ref{eqn:lorentzian}), and treated
the peak frequencies of the fitted Lorentzians as the QP frequencies.

\begin{figure}
    \centering
    \includegraphics[width=0.5\textwidth]{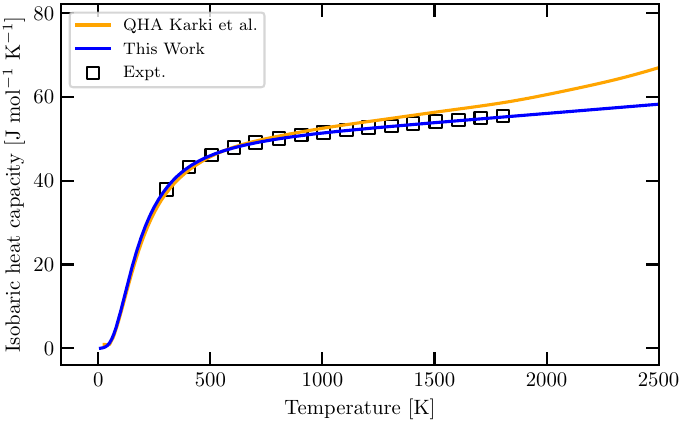}
    \caption{Isobaric heat capacity versus temperature. 
Symbols denote the measured values of Isaak {\em et al.}~\cite{Isaak_1989}, the 
orange line is the result of the DFPT QHA calculations by Karki et al.~\cite{Karki2000} 
and the blue line is the result of our calculations within the QPA.}
    \label{figd1}
\end{figure}

The comparison between our QPA calculation, the QHA calculations
of Karki {\em et al.}, and the measurements of Isaak, Anderson, and Goto are
presented in Fig.~\ref{figd1}. 
The QPA heat capacities are in excellent agreement with 
experiment at all values of $T$ for which experimental
data is available. Therefore the inaccuracy of the QHA calculation
at high $T$ is unlikely to be a result of treating
phonons as quasi-independent excitations. 
It must be caused by the QHA failing to account for the {\em self energies} 
of these phonons. It is their self energies that turn them from NMVs into phonon QPs~\cite{wallace,mattuck}.

\subsection{Thermal broadening of bands}
\label{section:broadening}
\begin{figure}[!]
    \includegraphics[width=0.5\textwidth]{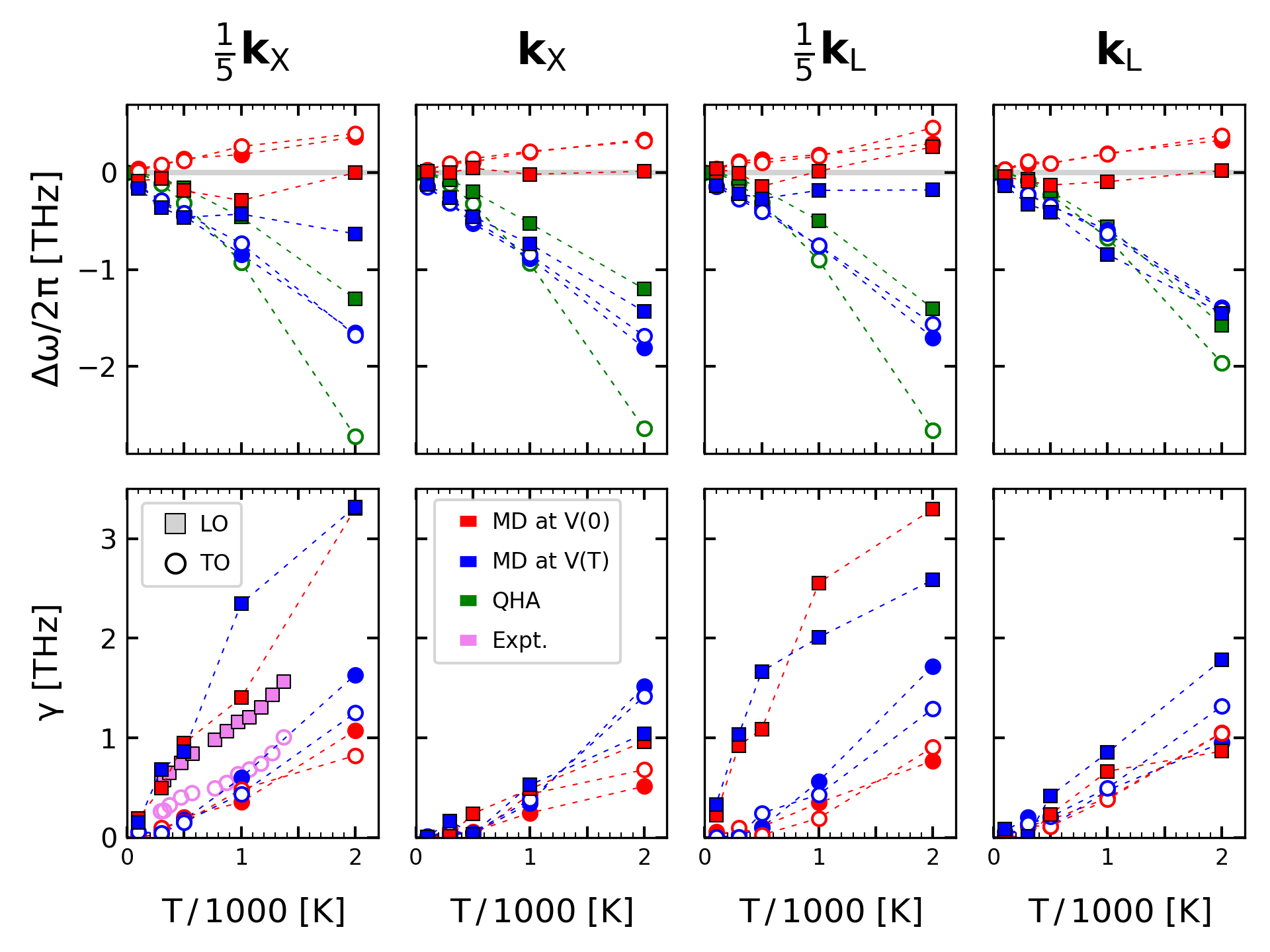}
    \caption{Results of fitting Lorentzian functions to optical mode spectra at, from left
to right, the wavevectors ${\frac{1}{5}\bk_X}$, ${\bk_X}$, ${\frac{1}{5}\bk_L}$, and ${\bk_L}$, 
where ${\bk_X\equiv\frac{1}{2}\left(\bb^1+\bb^3\right)}$, 
${\bk_L\equiv\frac{1}{2}\left(\bb^1+\bb^2+\bb^3\right)}$, and ${\{\bb^1,\bb^2,\bb^3\}}$
is the set of primitive reciprocal lattice vectors. The
upper panels plot the difference between the fitted Lorentzian's peak frequency 
and the \templim phonon frequency versus $T$.
The lower panels plot the fitted Lorentzians' full widths at half maximum
versus $T$ and the left-most plot includes a comparison to the values measured by
Calandrini {\em et al.} at a much smaller wavevector~\cite{Calandrini2021}.
The insets specify the meanings of the symbols and their colors
in all eight plots.}
    \label{fig:lorentzianshifts}
\end{figure}

\begin{figure}[!]
    \includegraphics[width=0.5\textwidth]{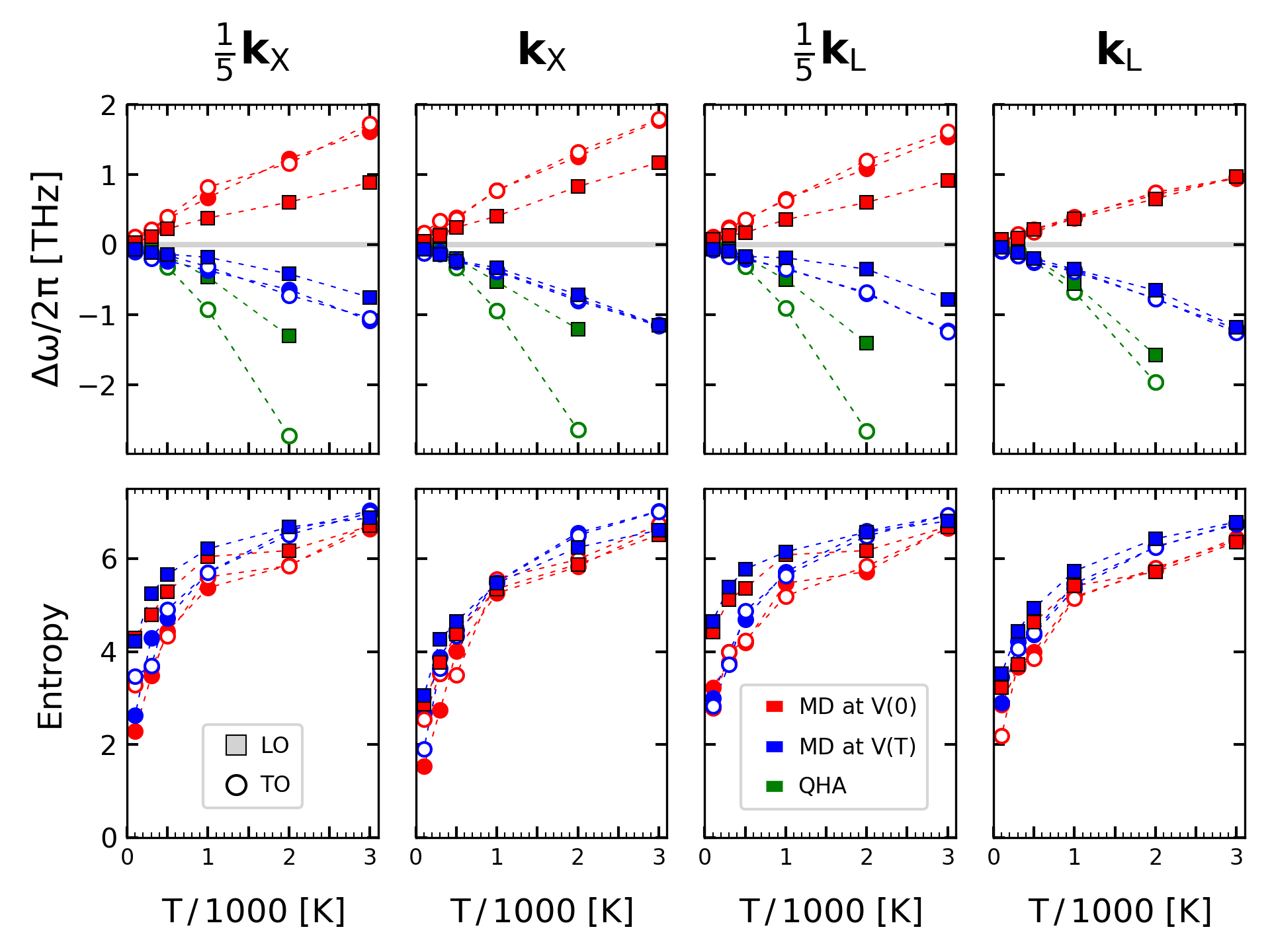}
    \caption{Upper panels: Difference between the mean of each optical mode's spectrum
and the mode's \templim~frequency as a function of $T$ in our ${V(0)}$ MD
simulations, our ${V(T)}$ MD simulations, and our QHA calculations. Lower panels: Entropies
of the optical mode spectra versus $T$. The insets specify
the meanings of the symbols and their colors in all eight plots.}
    \label{fig:opticshifts}
\end{figure}
In Sec.~\ref{section:deviations} we suggested some 
mechanisms by which LO bands can 
be broadened and shifted in frequency by acoustic modes,
without acoustic bands degrading to the same degree. 
In Sec.~\ref{section:thermal_expansion} we showed that thermal
expansion causes most bands to shift to lower frequencies 
as $T$ increases, but is not the cause of the anomalous
broadening of the LO band.
In this section we take a closer 
look at thermal broadening of bands.

\begin{figure}[!]
    \includegraphics[width=0.5\textwidth]{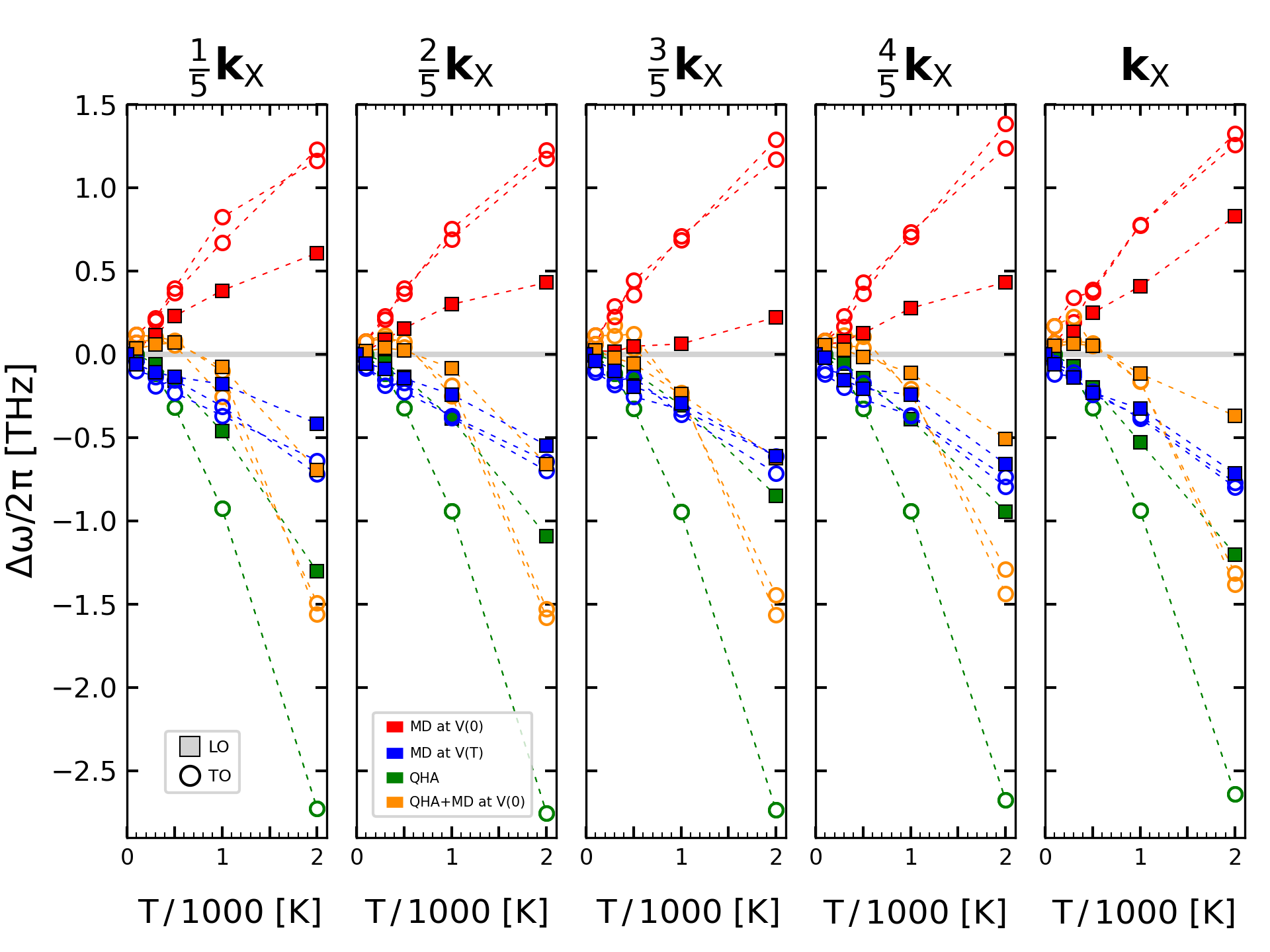}
    \caption{LO and TO frequency shifts as a function of $T$ at all finite wavevectors in our simulations
between $\Gamma$ and $X$. 
The insets specify the meanings of the symbols and their colors, which are the same
as in Fig.~\ref{fig:opticshifts} apart from the inclusion of
the sums of the ${V(0)}$ MD shifts and the QHA shifts. These are plotted in orange to demonstrate
that they differ from the ${V(T)}$ MD shifts. 
}
\label{fig:compareqha}
\end{figure}

Fig.~\ref{fig:lorentzianshifts}
presents the results of fitting Lorentzian functions (Eq.~\ref{eqn:lorentzian})
to individual mode spectra (Eq.~\ref{eqn:mode_spectrum}).
The upper panels are plots of the $T$-induced frequency shifts
(Eq.~\ref{eqn:bomega}), 
${\Delta\omega_{\bk\mu}(T)\equiv\bomega_{\bk\mu}(T)-\bomega_{\bk\mu}(0)}$,
of the three optical modes at each of the four wavevectors 
${\frac{1}{5}\bk_X}$, $\bk_X$, ${\frac{1}{5}\bk_L}$, and ${\bk_L}$, 
where ${\bomega_{\bk\mu}(0)}$ is
the frequency of mode ${\bk\mu}$ at zero pressure
in the \templim~limit. 
In our MD simulations ${\bomega_{\bk\mu}(0)}$ is the frequency of normal
mode ${\bk\mu}$ at the $V$ that minimizes the potential energy, whereas
in the QHA it is its frequency at the $V$ that minimizes the sum of the potential energy
and the zero-point energy of the phonon modes. 
Results from our ${V(T)}$ and ${V(0)}$  MD simulations 
are plotted in blue and red, respectively, 
and our QHA results are plotted in green.
In each case the frequency shifts of TO and LO modes are plotted as circles
and filled squares, respectively. 
We use the same symbols and colors to plot the Lorentzian
linewidths ${\gamma_{\bk\mu}(T)}$ in the lower panels, but instead
of comparing to QHA linewidths, which we did not calculate, 
we compare to infrared reflectivity data of Calandrini {\em et al.}~\cite{Calandrini2021}.
We find good agreement, despite comparing calculations at ${\bk=\frac{1}{5}\bk_X}$ to
measurements at much smaller wavevectors. 

Overall, it is less easy to see firm trends in 
the data presented in Fig.~\ref{fig:lorentzianshifts} 
than it is to see them in Fig.~\ref{fig:opticshifts}, 
which uses the same symbols and colors to plot
the shifts in the mean, ${\Delta\omega_{\bk\mu}(T)}$,  and the entropies, ${S_{\bk\mu}(T)}$, 
of the mode distributions, ${f_{\bk\mu}(\omega)}$. 
Therefore we include Fig.~\ref{fig:lorentzianshifts} for two reasons:
to compare our MD linewidths with measured linewidths, and to illustrate
why, for the purpose of investigating the effects of increasing $T$, 
we have chosen to {\em not} assume that mode spectra are approximately Lorentzian.
The entropy of a Lorentzian distribution is ${\sim \log \gamma_{\bk\mu}}$, so
Fig.~\ref{fig:lorentzianshifts}(b) might be clearer if its vertical axis had a logarithmic
scale, but
the entropy is a more general measure of uncertainty than either the width
of the best-fit Lorentzian or its logarithm, and using it does not entail making any
physical assumptions.

The QHA's overestimation of the rates of decrease of
optical mode frequencies with respect to $T$
is very clear in Fig.~\ref{fig:opticshifts}. 
Figure~\ref{fig:compareqha} demonstrates that
the frequency shifts from our ${V(T)}$ simulations
are not well approximated by adding
the shift caused by increasing $T$ at a fixed $V$ of ${V(0)}$ 
to the shift caused by changing $V$ at a fixed $T$ of zero.
The former is positive, which is consistent with the fact that
phonon frequencies increase under pressure~\cite{Karki2000, Ghose_2006}:
As quantified in Sec.~\ref{section:MD}, when $V$ is fixed, $P$ increases as $T$ increases.
Increasing $P$ makes it more difficult for individual primitive cells of the crystal to 
strain in order to accommodate optical mode displacements.
This elastic response would reduce the energy cost of the
optical displacements; therefore optical frequencies are increased
when this response mechanism is suppressed.
The orange symbols in Fig.~\ref{fig:compareqha}
would coincide with their blue counterparts if 
changes to $T$ and $V$ contributed frequency shifts 
whose origins were approximately independent.
The fact that they do not suggests that $T$ has other important
effects. This highlights, again, the importance of interactions
between phonons and the fact that each 
mode spectrum is not the spectrum of bare NMVs, but of QPs. 

\begin{figure}[!]
    \includegraphics[width=0.5\textwidth]{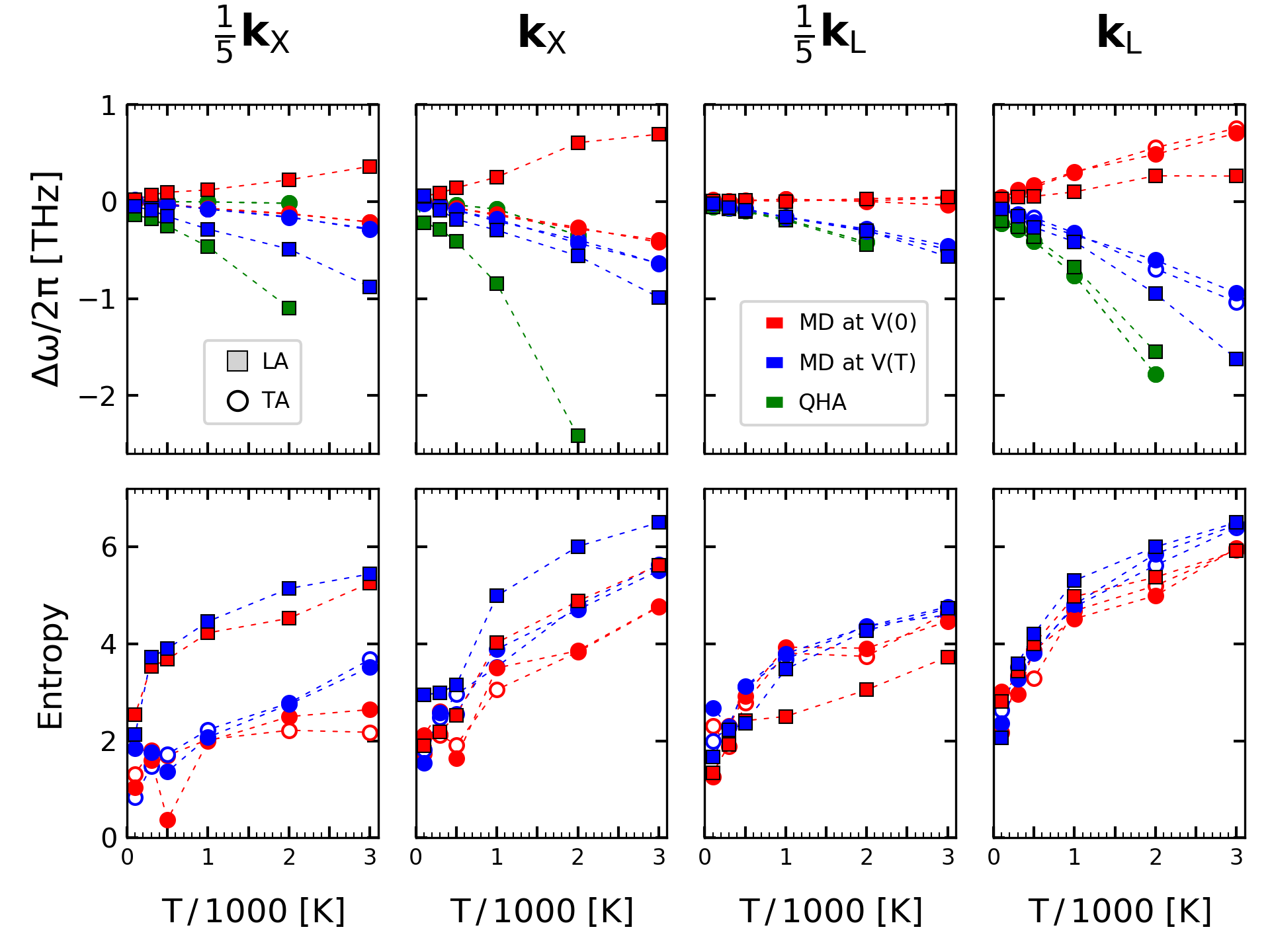}
    \caption{Upper panels: Difference between the mean of each acoustic mode's spectrum
and the mode's \templim~frequency as a function of $T$. Lower panels: Entropies
of the acoustic mode spectra versus $T$. The insets specify
the meanings of the symbols and their colors in all eight plots.}
    \label{fig:acousticshifts}
\end{figure}

Figure~\ref{fig:acousticshifts} shows that TA modes
are softened much less by $T$ than optical modes, 
especially at small wavevectors. At ${\bk=\frac{1}{5}\bk_X}$
the QHA shifts are negligible and the ${V(T)}$ and
${V(0)}$ shifts are almost equal. At ${\bk=\frac{1}{5}\bk_L}$, 
by contrast, the ${V(0)}$ shifts are negligible
and the ${V(T)}$ and QHA shifts are almost equal.
At all wavevectors except ${\bk=\frac{1}{5}\bk_L}$, 
the QHA overestimates the softening of the LA mode
at ${\SI{1000}{\kelvin}}$ and ${\SI{2000}{\kelvin}}$.
There is undoubtedly much to learn from a more detailed
analysis of the acoustic modes than we have performed.
The non-monotonicity of TA modes at low $T$ is particularly
intriguing.

Notice that, at $\bk_X$ and ${\bk_L}$, where acoustic
frequencies are as large as optical frequencies,
the entropies of acoustic and optical modes have similar magnitudes.
However near $\Gamma$ the entropies of acoustic modes, and particularly the TA modes,
are much smaller than those of optical modes.
Near $\Gamma$ is where acoustic frequencies are small enough 
that adiabatic decoupling is likely to provide acoustic
modes with the most protection from disorder
generated by the optical modes.

Also notice that, at all values of $T$,
the LA mode's entropy 
is significantly greater than the TA mode's entropy
at ${\bk=\frac{1}{5}\bk_X}$; whereas
at ${\bk=\frac{1}{5}\bk_L}$ it
is the TA mode whose entropy is larger at low $T$;
and at ${T\geq\SI{1000}{\kelvin}}$ 
the LA and TA modes have similar entropies.
Entropy quantifies uncertainty, which increases
with a mode distribution's width.
Therefore, Fig.~\ref{fig:acousticshifts} quantifies
our observation
in Sec.~\ref{section:acoustic_disruption} that, at low $T$,
there is broadening of the LA band along
${\Gamma\to X}$ and of the TA bands 
along ${\Gamma \to L}$.  

\begin{figure}[!]
    \includegraphics[width=0.5\textwidth]{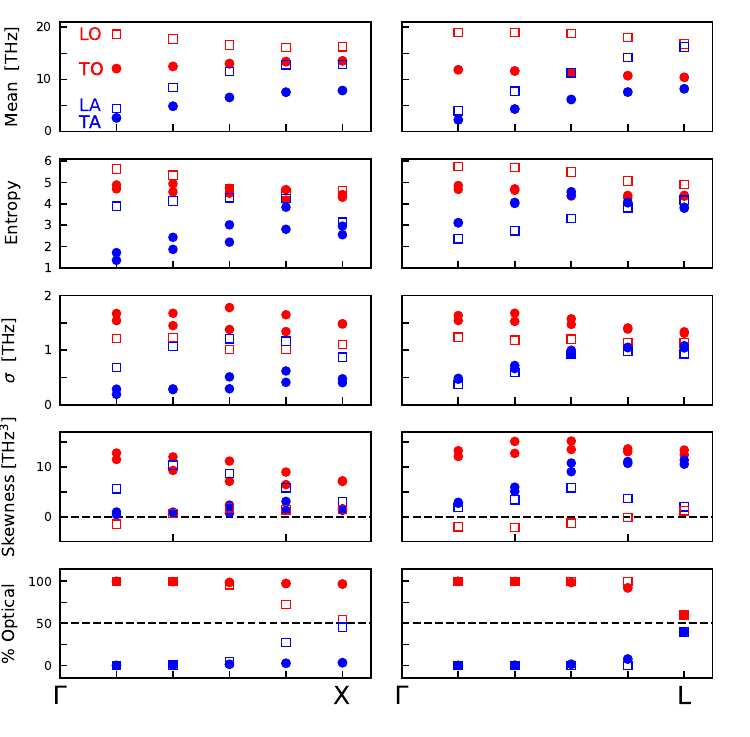}
    \caption{
Lowest panels: Degree to which each mode is an optical 
mode at ${T=0}$ as a function of $\bk$ along
wavevector paths ${\Gamma\to X}$ (left) and ${\Gamma\to L}$ (right).
Higher panels:
Characteristics of the normalized mode spectrum ${f_{\bk\mu}(\omega)}$ (Eq.~\ref{eqn:mode_spectrum})
as a function of wavevector ${\bk}$
at ${T=\SI{500}{\kelvin}}$. 
The four plotted characteristics of $f_{\bk\mu}$ are, 
from top to second-from-bottom, the mean ${\bar{\omega}_{\bk\mu}/2\pi}$, the entropy ${S_{\bk\mu}}$ (Eq.~\ref{eqn:entropy}), 
the standard deviation $\sigma$,
and the {\em skewness}, where $\sigma^2$ and the skewness are, respectively, the second (${m=2}$) and 
third (${m=3}$) central moments, ${\left(2\pi\right)^{-m}\sum_\omega\left(\omega-\bar{\omega}_{\bk\mu}\right)^mf_{\bk\mu}(\omega)}$.}
    \label{fig:peakdata}
\end{figure}

Figure~\ref{fig:peakdata} provides more insight into 
how the mode spectra change between 
the center and the boundary of the BZ.  It consists
of five vertically-stacked pairs of plots, with the
horizontal axes of each pair's left-hand and right-hand plots 
specifying one of the five finite wavevectors 
in our simulations along ${\Gamma\to X}$ and ${\Gamma\to L}$, respectively.
The data plotted in the lowermost pair of plots applies to the \templim limit, 
because it was calculated from normal mode eigenvectors.
We will discuss it below.
The four uppermost pairs are plots of
characteristics of the mode spectra from ${V(T)}$ MD simulations at ${T=\SI{500}{\kelvin}}$.
The uppermost plots are simply plots of the mean frequencies; i.e., the bands.
The next panels down are plots of the modes' 
entropies, ${S_{\bk\mu}}$ (Eq.~\ref{eqn:entropy}). The third panels down are plots
of the standard deviations ($\sigma$) of the mode spectra, which we include
only to allow $\sigma$ to be compared to $S_{\bk\mu}$ as a measure of thermal disorder.
The plots second from the bottom show that most of the
mode spectra are skewed towards higher frequencies, 
which would be consistent with them leaving behind 
a high-frequency tail when thermal expansions shifts
their means downwards.

We will not speculate further about why
most mode spectra have positive skewnesses.
However in Sec.~\ref{section:acoustic_disruption} we suggested
a mechanism that might contribute negatively to 
the skewness of only one of the six modes:
An acoustic perturbation that warps ${\lofield_\bk}$
is likely to reduce its magnitude, thereby lowering $\loangfreq_\bk$.
Therefore a thermal distribution of acoustic perturbations might contribute
low-frequency tails to LO mode spectra.
This might explain why the LO mode has a skewness that 
is smaller in magnitude than other modes when it is positive, 
why it is the only mode whose skewness is negative at some wavevectors, 
and why the skewness changes from negative to positive
between the BZ center and its boundary.

Now let us examine the plots of entropy in Fig.~\ref{fig:peakdata}.
Notice that, starting at  ${\bk=\frac{1}{5}\bk_X}$ or ${\bk=\frac{1}{5}\bk_L}$, 
the acoustic bands' entropies increase, initially,
as ${\abs{\bk}}$ increases.
This is to be expected because each mode's entropy vanishes in the \templim~limit 
and it is only the interactions between modes that makes it finite.
Acoustic modes' frequencies 
move closer to those of optical modes
as their wavevectors move away from the BZ center.
Therefore, we should expect their interactions
with optical modes to strengthen as ${\abs{\bk}}$ increases,
and this would increase their entropies.

If a mode was damped by many weak
interactions, its entropy would increase gradually and monotonically as either the number of interactions
or their strengths increased gradually.
However, sudden reductions of entropy occur to the LA mode and the TA modes along ${\Gamma\to X}$; and
there is a less dramatic reduction of the TA modes' entropies along ${\Gamma\to L}$. 
After increasing monotonically between ${\bk=\frac{1}{5}\bk_X}$ 
and ${\bk=\frac{4}{5}\bk_X}$, 
the entropies of the LA band
and the higher-entropy TA mode drop abruptly at ${\bk=\bk_X}$.
In the right-hand plot the entropies of both TA bands suddenly 
begin to decrease at ${\bk=\frac{4}{5}\bk_L}$.

These kinks in the entropy as a function of ${\bk}$ 
are interesting because they suggest the existence of
strong interactions between modes, which contribute
sizeable fractions of their total entropies.
If a mode's entropy 
only had contributions
from weak interactions, a sudden decrease
would require 
many interactions to weaken or vanish simultaneously (at the same wavevector).
This is improbable. Each kink is much more likely to be explained by the disappearance or weakening
of few very strong interactions, than by many weak ones.
Therefore a kink in a band's entropy is likely
to mean that, on the higher-entropy side of the kink, the mode
is involved in at least one particularly-strong interaction.

The fact that the entropy of the LA mode at ${\bk=\frac{1}{5}\bk_X}$ is
so high adds to the suspicion that it is caused by coupling
to the LO mode, because ${\laangfreq_\bk}$ is so low that 
only TA modes have similar frequencies.
However, the TA modes' entropies
are smaller, suggesting that they do not interact with other modes as strongly
as the LA mode does. Furthermore, if the high entropy of the LA band
was caused by almost-resonant coupling between LA and TA modes, it should decrease with increasing 
${\abs{\bk}}$, because ${\laangfreq_\bk}$ and ${\taangfreq_\bk}$ rapidly diverge.
For a similar reason, the fact that the TO modes' entropies remain relatively constant
as ${\laangfreq_\bk}$ approaches ${\toangfreq_\bk}$
makes it unlikely that the LA entropy has a large contribution from LA-TO coupling. 
The LA-TO interaction should strengthen as their frequencies 
get closer, which would increase the TO entropy.

However, if the LA entropy was
high due to LA-LO coupling, we would not necessarily see the LO mode
entropy increase as ${\laangfreq_\bk}$ increased.
If the high LO entropy was caused by the LA mode 
warping ${\lofield_\bk}$, the effects of this
disruption on the LO band would not necessarily increase as ${\laangfreq_\bk}$ drew
closer to ${\loangfreq_\bk}$, because the strength of this coupling mechanism 
does not rely on the coupled modes having similar frequencies.
Furthermore, the degree to which ${\lofield_\bk}$ shifts
${\loangfreq_\bk}$ upwards decreases with increasing wavevector. Therefore,
if the high LO entropy was caused by warping of ${\lofield_\bk}$, it
should decrease with increasing ${\abs{\bk}}$, as it can be seen
to do in Fig.~\ref{fig:peakdata}.
This diminishing importance of ${\lofield_\bk}$ as the BZ boundary
is approached would also contribute negatively to the LA band's entropy, but
the increase of the LA band's frequency means that its 
interactions with most optical modes strengthen with increasing
wavevector as it gets closer to resonance with them.
This positive contribution to its entropy might 
more than offset the entropy reduction 
caused by its interaction with the LO mode weakening.
A further clue that the high entropy of the LA mode
is caused by its interaction with the LO mode via
warping  of ${\lofield_\bk}$ is that
${\lofield_\bk}$ vanishes at $X$, which 
is where the LA entropy suddenly drops. The value
it drops to is similar to that of the blue TA mode, 
which increases monotonically
between ${\Gamma}$ and ${X}$.

An obvious question to address is 
why the entropy of one of the
TA bands increases more than the other, before
also dropping suddenly at ${\bk=\bk_X}$.
The entropy of this 
band increases much more rapidly than the
entropy of the other TA band, despite their mean
frequencies remaining equal. However, 
it is important to note that the eigenvectors
of two degenerate modes can be rotated in 
the two-dimensional space that they span, without
changing their frequencies. For this reason, 
the two TA bands do not have independent identities
and, for some rotations of TA eigenvectors within
the subspace that they span, the entropies
of the bands would be equal.
Therefore we should not refer to entropies of the
two TA bands individually; we should say that
the combined entropy
of the TA modes increases with
increasing wavevector along ${\Gamma\to X}$, 
before suddenly  decreasing at ${\bk=\bk_X}$.
This means that the results plotted in the left-hand panels of
Fig.~\ref{fig:peakdata} are consistent with 
acoustic warping of ${\lofield_\bk}$, which mostly
involves the LA band
near $\Gamma$, but with
the TA modes becoming more involved as 
the BZ boundary is approached.

The sudden decrease of the TA entropy at ${\bk=\bk_X}$ 
is difficult to explain if its increase between ${\bk=\frac{1}{5}\bk_X}$
and ${\bk=\frac{4}{5}\bk_X}$ is caused by coupling to the TO modes.
The data plotted in the bottom left panel of Fig.~\ref{fig:peakdata}
also makes TO-TA coupling less likely to be responsible for 
the TA entropy being nonmonotonic.
This is a plot of the degree to which each mode is
an optical mode.
An optical mode modulates the cation-anion displacement, 
whereas an acoustic mode modulates their center of mass. Therefore, instead
of expressing eigenvectors in terms of displacements of Mg and O ions, 
we can express them in terms of Mg-O relative displacements
and center of mass displacements. The former 
vanishes for an acoustic mode, the latter vanishes
for an optical mode. Therefore, we use $100$ times the 
magnitude of latter divided by the magnitude of
their sum to quantify the degree to which each
mode is an optical mode.
What we find is that, at ${\bk=\frac{4}{5}\bk_X}$,
the LA and LO modes become significantly optical
and acoustic, respectively. In other words, 
they mix strongly to become acoustic/optical hybrids.

At ${\bk=\bk_X}$ the polarization vectors of the LA and LO modes
are identical, by symmetry. Mathematically, the difference
in their frequencies can be viewed as a consequence
of their normalized eigenvectors being different scalar
multiples of the same polarization vector. 
The LO and LA eigenvectors are the polarization vector multiplied by the square
roots of the O and Mg masses, respectively.

It is well known that when two modes couple strongly enough, they tend to hybridize: 
each one's eigenvector becomes a mix of both eigenvectors,
such that the modes described by the hybrid eigenvectors
interact much more weakly. This is what causes avoided
crossings in band structures.
The LO-LA hybridization does not tell us anything directly 
about TO-TA coupling, but the bottom left plot
of Fig.~\ref{fig:peakdata} shows that
the TO and TA modes do not hybridize noticeably.
At ${\bk=\frac{3}{5}\bk_X}$ each TA (TO) mode
is only ${1.4\%}$ optical (acoustic), and 
this increases to only ${3.2\%}$ at ${X}$.
Therefore the sudden decrease of the TA modes' entropy
at ${\bk=\bk_X}$ is unlikely to be a consequence of hybridization and we must
look elsewhere for an explanation of it.
The involvement of the TA modes in the acoustic-LO coupling
is an obvious alternative explanation.

Now let us turn our attention to the path ${\Gamma\to L}$.
The bottom-right plot
of Fig.~\ref{fig:peakdata} shows that the TO and TA
modes begin to hybridize at ${\bk=\frac{4}{5}\bk_L}$, and 
that all modes become 
hybrids of
optical and acoustic modes at
${\bk=\bk_L}$, with each mode's optical/acoustic ratio being
approximately ${60/40}$ or ${40/60}$.
Therefore the non-monotonic behaviour of the TA entropy
might be explained by its sudden decoupling from the TO mode
when they transform into TO/TA hybrids.
This would make sense because they are both transverse modes, which
become very close in frequency near ${\bk=\bk_L}$.
Therefore it seems plausible to interpret the kink in the entropy as
a signature of strong TO-TA coupling at ${\bk<\frac{4}{5}\bk_L}$, 
which would explain why the entropy 
of the TA mode is greater than that of the LA mode near $\Gamma$.

There are also some strong hints that acoustic
modes are disrupting ${\lofield_\bk}$ along ${\Gamma\to L}$.
Firstly, the LO mode has the highest entropy, despite
being the least active mode at low $T$.
Secondly, the LO mode has a negative skewness until 
${\lofield_\bk}$ vanishes at ${\bk=\bk_L}$, which is consistent
with the acoustic warping of ${\lofield_\bk}$ giving 
it a low-frequency tail. Thirdly, the variation
of the skewness with $\bk$ appears to mirror
the variation of the mean frequency: As one increases
the other decreases.
This is consistent with the negative LO skewness 
being caused by this mechanism because it is
the increasing strength of ${\lofield_\bk}$ that
causes ${\loangfreq_\bk}$ to increase between
$L$ and $\Gamma$. As the field's
contribution to ${\loangfreq_\bk}$ increases in magnitude,
${\loangfreq_\bk}$ becomes more vulnerable to being lowered
by acoustic perturbations.

All of the acoustic-LO coupling might be LA-LO coupling, 
but it is also possible that the TA mode is involved. 
However, near the BZ center, strong TA-LO coupling via TA disruption of ${\lofield_\bk}$
requires the lateral extent of the LO phonon to be not much greater
than its wavelength.  This is because, if we approximately describe
${\brhobk}$ as a stack of parallel {\em uniformly}-charged infinite planes,
a transverse displacement of each plane would not
alter the constant field emanating from it. 
At finite $T$ the region perturbed by an LO phonon in a plane
perpendicular to $\bk$ has a finite area. Therefore
transverse displacements 
would change ${\lofield_\bk}$ near the boundary of this region
and, if a large enough fraction of the atoms participating
in the LO motion were affected by this change, 
a TA phonon could couple significantly to it by
this mechanism. 

When the LO wavelength is small, on the other hand, the 
crystal planes displaced laterally by a TA phonon 
cannot be treated as uniformly charged; their atomistic structures have a significant influence on the value of ${\lofield_\bk}$.
Therefore, near the BZ boundary there could be strong TA-LO coupling for
LO phonons whose lateral extents are much larger than their wavelengths.

\section{Conclusions}
\label{section:conclusions}
In this work we have studied strong phononic correlation in crystalline MgO with a method
of calculating vibrational spectra
that does not suffer from the limitations of perturbation theories or mean-field theories,
and which can produce accurate wavevector-resolved spectra with enough detail to observe bandstructures.
These calculations, and recent similar calculations by Lahnsteiner and Bokhdam~\cite{lahnsteiner}, have been made possible by force fields whose parameters
are fit to vast quantities of data calculated {\em ab initio}~\cite{Ercolessi_1994,Tangney_Scandolo_2003, behler_2007,Csanyi_PRL_2010,machine_learning_2021}, and whose
low computational expense, relative to {\em ab initio} MD, makes
calculations of {\em high resolution} spectra possible.

As shown in Appendix~A and Ref.~\onlinecite{theory_paper},
the method we have used to calculate spectra
from our MD simulations is much more powerful and generally-applicable than
existing derivations of it suggest~\cite{dove_1993,vanderbilt_ice,Ladd_1986,Sun_2010,Zhang2014,Sun_2014,Meyer_2011,Kirchner_2013,tuckerman,meunier}.
It is not a new method, having been used to calculate frequency- and wavevector-resolved
spectra at least as far back as 2006~\cite{Tangney2006}, and to calculate spectra
as functions of frequency only for many years prior to that. 
However existing derivations of it 
do not justify using it to calculate spectra 
at very high $T$, when phononic correlation is strong, as we have done.
Therefore it has usually been interpreted as a method of
calculating the VDOS at thermal equilibrium when phonons do not
interact (i.e., in the \templim~limit) or when their
interactions are sufficiently simple that each phonon is
an exponentially-decaying harmonic wave, in which case its spectral peak is Lorentzian~\cite{Koker_2009,Sun_2010,Sun_2014,lahnsteiner}.

However, we have shown that it is an exact method, whose applicability is not restricted to crystals, to low $T$, 
or even to states of thermal equilibrium.
This means that it is a powerful tool for studying vibrations in materials, which complements
and is complemented by, perturbation theories. It can also be used to assess the accuracies
of perturbative methods, mean-field methods, and other methods that rely on simplifying assumptions, such as 
the assumption of a state of thermal equilbrium~\cite{Monacelli_2021}.
For these reasons, 
it seems likely to play an increasingly important
role as it becomes possible to calculate forces accurately and efficiently for 
more materials. 

For example, 
it should help with the development
of materials that are either transparent or opaque to THz
radiation in a particular frequency window. 
In other words, it should help to guide us in the 
rational design of materials that behave as THz bandpass filters
for use as active or passive components of THz devices.

Another application is in the study of order-disorder transitions
between crystalline phases, which occur when a single phonon band melts 
but the other bands do not, causing displacements and velocities
along the eigenvectors of the melted band's modes to become disordered.
This critical weakening of correlations 
causes both the time average and the spatial average
of the displacements to vanish. As a result,
when the crystal is observed at either a low spatial resolution
or a low temporal resolution, its symmetry appears to have increased.
The ability to see how the vibrational spectrum changes as $T$ crosses
the transition temperature may lead to step changes
in our understandings of these transitions, or even to a step
change in our understanding of this class of transitions.

A third application is in the study of materials at very high $T$,
such as thermal barrier coatings, where there are high
densities of defects, and therefore high densities
of vibrations that are localized in spacetime and
delocalized in reciprocal spacetime.
A fourth application, which is related to the third, is in the identification or design of materials with
reduced or enhanced thermal conductivities. For example, the thermal conductivities
of materials with low-frequency localized {\em rattler} modes~\cite{Christensen_2008} are low because
the energy of these modes, being localized in spacetime, is delocalized
in \wk-space. Therefore they overlap with acoustic modes in reciprocal spacetime,
which allows the acoustic modes to couple with them, and with one another through them,
to disperse their energy.

However the generality of this method means that one of the
most important applications of it may be to study general properties
of strong phononic correlation.
The simplest and most obvious first step is to study the strengthening of phononic
correlation in a simple prototype crystal as it is heated. This has been the purpose of this work.

We have found that, despite the apparent banality of crystalline MgO at thermal equilibrium, 
there is so much to be learned from the spectra presented herein that we have been forced to focus most of our 
analyses of them on a few of their gross features. We provide our raw ${V(T)}$ spectra, in the form of tabulations
of ${\hE^\K(\bk,\omega)}$ at each value of $T$, as supplementary material in case others wish to 
analyse it further or differently~\cite{supplementary}.

By calculating spectra at both constant volume
and constant pressure, we have been able to show conclusively that some
features of the spectrum's $T$ dependence, 
such as the tendency of bands to soften as
$T$ increases,
are consequences of thermal expansion; and 
that others, such as the rapid degradation of the LO band, 
are not. We have shed substantial light on the strengths and limitations
of the quasiharmonic approximation and we have shown that some
of its weaknesses are rectified by the quasiparticle approximation.

Most importantly, we have identified, and discussed in some detail, two physical
considerations that are likely to have observable consequences for many or most
crystals. The first is that LO phonons are highly sensitive
to the amplitudes of acoustic phonons because a large
fraction (${\sim\frac{1}{2}}$ for MgO) of ${\loangfreq_\bk}$ is contributed by the
LO mode's intrinsic field, which acoustic phonons warp. This
simple mechanism must occur, to some degree, in every partially-ionic crystal, because
every LO phonon has an intrinsic field that opposes its motion and increases its frequency.
The second consideration is that, despite this strong acoustic-LO coupling,
long wavelength acoustic phonons are adiabatically
decoupled from LO modes: they strongly modulate the energies
and frequencies of LO modes, while maintaining approximately-constant
energies and frequencies themselves.

The feature of the spectrum's $T$-dependence that led us to these ideas and
explanations is very similar to features that have been observed experimentally in 
the spectra of other ionic crystals since the 1960s~\cite{Woods1960, Woods1963, Raunio_1969, Raunio_1969_2, Lowndes_1970, Chang_1972, Nilsson_1981, Cowley_1983, Shen_2020, Shen_2021}. It has also been
studied in MgO  and similar materials using perturbation theory~\cite{Giura2019, Calandrini2021, Togo_2022}, 
but it has lacked a simple and intuitive physical explanation until now.

\section*{Acknowledgments}
GC was supported through a studentship in the Centre for Doctoral Training on Theory and Simulation of Materials at Imperial College London funded by the EPSRC (EP/L015579/1).

\appendix
\section{Distribution of the kinetic energy of a vibrating string in reciprocal spacetime}
\label{section:string}
Consider a taut string, of uniform mass per unit length $\rho$,
whose ends are fixed at ${x=-L/2}$ and ${x=L/2}$,
and which lies along the $x$ axis when it is at equilibrium. 
Its displacement from equilibrium at 
spacetime coordinates ${(x,t)}$ 
is denoted by ${u(x,t)}$, 
and its kinetic energy per unit length
is ${\kappa(x,t)\equiv\frac{1}{2}v(x,t)^2}$, where ${v\equiv\sqrt{\rho\dd{x}}\,\dot{u}}$
and
${\dd{x}}$ is an infinitesimal length. 

Let us assume that the string only moves during the time
interval ${(-\tmax/2,\tmax/2)}$, or that it is only observed
for times ${t\in(-\tmax/2,\tmax/2)}$.
In the latter case, we assume that, at each $x$, ${v(x,t)}$ has been
tapered smoothly, but arbitrarily rapidly, to zero at
times ${-\tmax/2}$ and ${\tmax/2}$.
Similarly, ${v(x,t)}$ is both smooth and vanishes for ${x\notin (-L/2,L/2)}$. 
Therefore, for all of our purposes within this appendix,  
${\int_{-\tmax/2}^{\tmax/2} \dd{t}}$ is equivalent to ${\int_\realone\dd{t}}$
and ${\int_{-L/2}^{L/2} \dd{x}}$ is equivalent to ${\int_\realone\dd{x}}$.

The average, over interval ${(-\tmax/2,\tmax/2)}$, of the string's
kinetic energy per unit length at $x$ is
\begin{align}
\expval{\kappa}_t(x) 
&= \frac{1}{2\tmax}\int_\realone v(x,t)^2\dd{t}
\label{eqn:kappa}
\\
&= \frac{1}{2\tmax}\int_\realone \brv^{*}(x,\omega)\brv(x,\omega)\dd{\omega},
\label{eqn:parseval1}
\end{align}
where ${\brv(x,\omega)\equiv \left(2\pi\right)^{-1/2}\int_\realone v(x,t)e^{-i\omega t}\dd{t}}$ is
the unitary Fourier transform of ${v(x,t)}$ with respect to time, and we have
used {\em Parseval's theorem}~\cite{strichartz} to reach Eq.~\ref{eqn:parseval1}
from Eq.~\ref{eqn:kappa}.
Since ${v(x,t)}$ is real, ${\brv^*(x,\omega)=\brv(x,-\omega)}$, which implies
that ${\brv^{*}(x,\omega)\brv(x,\omega)=\brv(x,-\omega)\brv(x,\omega)=\brv^{*}(x,-\omega)\brv(x,-\omega)}$.
Therefore, ${\frac{1}{2}\int_\realone \dd{\omega} \brv^{*}(x,\omega)\brv(x,\omega) = \int_{\realpos}\dd{\omega}\brv^{*}(x,\omega)\brv(x,\omega)}$.
If we integrate ${\expval{\kappa}_t(x)}$ over all $x$ and use Parseval's theorem again we find that the
average kinetic energy of the entire string is
\begin{align*}
\expval{\K} 
& =   \frac{1}{\tmax}\int_{\realpos}\dd{\omega}\int_{\realone}\dd{k}\tv^*(k,\omega)\tv(k,\omega)
\end{align*}
where ${\tv(k,\omega)}$ is the unitary FT of ${v(x,t)}$ with respect to both $x$ and $t$, i.e., 
\begin{align*}
\tv(k,\omega)& =\frac{1}{2\pi}\int_\realone \dd{\omega} e^{-i\omega t}\int_\realone\dd{k} e^{-ikx} v(x,t)
\\
              &  = \left(\frac{1}{2\pi}\int_\realone \dd{\omega} e^{i\omega t}\int_\realone\dd{k} e^{ikx} v(x,t)\right)^* 
\\
&= \tv^*(-k,-\omega)
\end{align*}
Therefore, the string's average kinetic energy divided by its length is
\begin{align}
\frac{\expval{\K}}{L} = \frac{1}{L\tmax}\int_{\realpos}\dd{\omega}\int_\realone\dd{k}\tv(-k,-\omega)\tv(k,\omega).
\label{eqn:kinetic2}
\end{align}

The VVCF is defined as
\begin{align*}
C(x,t) 
&\equiv \expval{\expval{v(x_0,t_0)v(x_0+x,t_0+t)}_{x_0}}_{t_0}
\\
&=\frac{1}{L\tmax}\int_{\realone}\dd{t_0}\int_\realone\dd{x_0} v(x_0,t_0)v(x_0+x,t_0+t)
\end{align*}
If we express ${v(x_0,t_0)}$ and ${v(x_0+x,t_0+t)}$
in terms of $\tv$, perform the integrals over $x_0$ and $t_0$ before
the integrals over frequencies and wavevectors, and use the identities
${2\pi\delta(k)=\int_{\realone}e^{ikx_0}\dd{x_0}}$ and
${2\pi\delta(\omega)=\int_{\realone}e^{i\omega t_0}\dd{t_0}}$, this becomes
\begin{align*}
C(x,t) = \frac{1}{L\tmax} \int_{\realone} \dd{\omega}e^{i\omega t}\int_{\realone}\dd{k} e^{ikx}\tv(-k,-\omega)\tv(k,\omega)
\end{align*}
Therefore the FT of ${C(x,t)}$ with respect to both $x$ and $t$ is
${\tC(k,\omega) = \left(2\pi/L\tmax\right)\tv(-k,-\omega)\tv(k,\omega)}$, and Eq.~\ref{eqn:kinetic2}
becomes
\begin{align}
\frac{\expval{\K}}{L} = \frac{1}{2\pi}\int_{\realpos}\dd{\omega}\int_\realone\dd{k}\tC(k,\omega).
\label{eqn:kefinal1}
\end{align}
Now let us assume that string's velocity is sampled at $N_t$ evenly-spaced
times between ${-\tmax/2}$ and ${\tmax/2}$ and at $N_x$ evenly-space points between
${-L/2}$ and ${L/2}$. 
Then the VVCF becomes
\begin{align*}
C(x,t) = \frac{1}{N_tN_x} \sum_{t_0}\sum_{x_0} v(x_0,t_0)v(x_0+x,t_0+t), 
\end{align*}
where the sums are over the sampled times and positions.

Now, because ${v(x,t)}$ vanishes when 
${x\notin(-L/2,L/2)}$, the smallest observable wavevector and the smallest
observable difference between two wavevectors are both equal to
${\Delta k \equiv \pi/L}$. Similarly, because ${v(x,t)}$ vanishes when 
${t\notin(-\tmax/2,\tmax/2)}$, the smallest observable frequency 
and the smallest observable difference between two frequencies are both equal to ${\Delta\omega\equiv \pi/\tmax}$.
This implies that there are only $N_t$ observable frequencies and $N_x$ observable wavevectors, 
and the integrals in Eq.~\ref{eqn:kefinal1} must be approximated as sums over these observable values.
If we define 
${\hC(k,\omega)\equiv \tC(k,\omega)\Delta\omega\Delta k/2\pi}$, 
we can express the discretized version of
Eq.~\ref{eqn:kefinal1} as
\begin{align*}
\frac{\expval{\K}}{L} = \sum_{\omega>0}\sum_{k}\hC(k,\omega),
\end{align*}
and the discrete FT of ${C(x,t)}$ as
\begin{align*}
\hC(k,\omega) = \frac{1}{N_tN_x}\sum_t\sum_x C(x,t) e^{-ikx}e^{-i\omega t}.
\end{align*}
Therefore the average kinetic energy of the string per unit of its length
is a sum over all sampled frequencies and wavevectors of the discrete FT
of the VVCF. The linearity of the FT implies that each term ${\hC(k,\omega)}$ in this sum is the contribution
to the average kinetic energy of motions with wavevector $k$ and frequency $\omega$.
\section{The force field} \label{MDFF}
We use a classical dipole-polarizable potential of the form described in Refs.~\cite{Tangney_Scandolo_2002_2},~\cite{Tangney_TiO2_2010}, 
and~\cite{Sarsam_Finnis_Tangney_2013}. It includes a sum of purely-pairwise interaction energies of the form
\begin{align*}
    U_{ij}(r) = \frac{q_i q_j}{r_{ij}} + D_{ij}[e^{\gamma_{ij} [1 - (r_{ij}/r^0_{ij})]} - 2 e^{(\gamma_{ij}/2) [1-(r_{ij}/r^0_{ij})]}],
\end{align*}
where $r$ is the distance between ions $i$ and $j$, and
$q_i$ and $q_j$ are their charges. The parameters $D_{ij}$, $\gamma_{ij}$ and $r^0_{ij}$ define the 
{\em Morse potential} and, along with the charges, are among the parameters that are fit to  reproduce
forces, stress tensors, and energy differences calculated {\em ab initio} with density functional theory.
A list of all potential parameters and their values are provide in Table~\ref{tab}.

In addition to the pairwise part of the potential, the oxygen anion is assigned a polarizability, $\alpha$, 
which is used to assign a dipole moment $\mathbf{p}_i$ to each anion as follows:
\begin{align*}
\mathbf{p}_i = \mathbf{p}_i^{SR} + \alpha \mathbf{E}_i(\{\mathbf{p}_i\})
\end{align*}
where the second term on the right hand side is the dipole moment induced on
ion $i$ by the local electric field from the charges and dipole moments 
of all other ions, and
\begin{align*}
    \mathbf{p}_i^{SR} &= \alpha \sum_{j \neq i} \frac{q_{ij} \mathbf{r}_{ij}}{r^3_{ij}} f_{ij}(r_{ij})
\intertext{where}
    f_{ij}(r_{ij}) &= c_{ij} \sum_{k=0}^4 \frac{(b_{ij} r_{ij})^k}{k!} e^{-b_{ij} r_{ij}}.
\end{align*}
is a short range contribution to the dipole from an anion's electron cloud changing shape as its neighbouring ions move.
Since each dipole moment depends on the values of all others, the set of all dipole moments is calculated, at each
set of ionic positions, by iterating them to self-consistency~\cite{Tangney_Scandolo_2002_2}.
After self consistency has been achieved 
charge-dipole and interaction energies are added to ${U_{ij}}$ for each pair of ions ${(i,j)}$ 
that includes an oxygen ion and dipole-dipole interaction energies are added to it when $i$ 
and $j$ are both oxygen ions.

\begin{table}[h!]
\begin{tabular}{c c c c c c} 
 \hline
  & Mg & O & Mg-Mg & Mg-O & O-O \\
 \hline\hline
 charges &&&&& \\

 q & 1.415382 & -1.415382 & & &\\ 
 \hline
 $ D $ & & & 0 & 3.0945 $\times$ 10$^{-3}$  & 2.161$\times$ 10$^{-3}$\\
 \hline
 $\gamma$ &  &  & 25.1073 & 10.1621 & 8.9925\\
 \hline
 $r^0$ & & & 26.9788 & 4.9640 & 6.1491 \\
 \hline \hline 
 dipoles &&&&& \\
 $\alpha$ & 0 & 9.6565 &  &  & \\
 \hline
 $b$ & & & 0 & 1.8713 & 0 \\
 \hline
 $c$ & & & 0 & -1.6809 & 0 \\
\end{tabular}
\caption{Force field parameters in atomic units.}
\label{tab}
\end{table}


%
\end{document}